\documentclass[a4paper,fleqn,usenatbib]{mnras}

\usepackage[T1]{fontenc}
\usepackage{ae,aecompl}

\usepackage{graphicx}	
\usepackage{amsmath}	
\usepackage{amssymb}	
\usepackage{natbib}
\usepackage{fancyhdr}
\usepackage{hyperref}

\title[Efficiency of the \emph{Pristine} survey]{The \emph{Pristine} survey III: Spectroscopic confirmation of an efficient search for extremely metal-poor stars\thanks{This paper is based on photometric data obtained with CFHT programs 15AC20, 15AF14, 15AF97, 16AC20, 16AC98, and 16AF14 and spectroscopic data from INT and WHT programs C71 and N5 in semester 2016A.}}

\author[K. Youakim et al.]{
K. Youakim$^{1}$\thanks{E-mail: kyouakim@aip.de},
E. Starkenburg$^{1}$,
D. S. Aguado$^{2,3}$,
N. F. Martin$^{4,5}$,
M. Fouesneau$^5$,
\newauthor
J. I. González Hernández$^{2,3}$, 
C. Allende Prieto$^{2,3}$,
P. Bonifacio$^{6}$,
M. Gentile$^{7}$,
\newauthor
C. Kielty$^{8}$,
P. C\^ot\'e$^{9}$,
P. Jablonka$^{7}$,
A. McConnachie$^{9}$,
R. S\'anchez Janssen$^{10}$,
\newauthor
E. Tolstoy$^{11}$,
and K. Venn$^{8}$
\\
$^{1}$Leibniz-Institut f\"ur Astrophysik Potsdam, An der Sternwarte 16, Potsdam 14482, Germany\\
$^{2}$Instituto de Astrof\'isica de Canarias, Vía Láctea, 38205 La Laguna, Tenerife, Spain\\
$^{3}$Universidad de La Laguna, Departamento de Astrof\'isica, 38206 La Laguna, Tenerife, Spain\\
$^{4}$Universit\'e de Strasbourg, CNRS, Observatoire astronomique de Strasbourg, UMR 7550, F-67000 Strasbourg, France\\
$^{5}$Max-Planck-Institut f\"ur Astronomie, K\"onigstuhl 17, D-69117 Heidelberg, Germany\\
$^{6}$GEPI, Observatoire de Paris, PSL Research University, CNRS, Place Jules Janssen, \\
92190 Meudon, France\\
$^{7}$Laboratoire d'Astrophysique, \'Ecole Polytechnique F\'ed\'erale de Lausanne (EPFL) Observatoire de Sauverny, CH-1290 Versoix, \\
Switzerland\\
$^{8}$Dept. of Physics and Astronomy, University of Victoria, P.O. Box 3055, STN CSC, Victoria BC V8W 3P6, Canada\\
$^{9}$National Research Council, Herzberg Astronomy \& Astrophysics, 5071 West Saanich Road, Victoria, BC, V9E 2E7, Canada\\
$^{10}$UK Astronomy Technology Centre, Royal Observatory, Blackford Hill, Edinburgh EH9 3HJ, UK\\
$^{11}$Kapteyn Astronomical Institute, University of Groningen, Landleven 12, 9747AD Groningen, Netherlands\\
}

\date{Accepted XXX. Received YYY; in original form ZZZ}

\pubyear{2017}

\begin{document}
\label{firstpage}
\maketitle

\begin{abstract}
The \emph{Pristine} survey is a narrow-band, photometric survey focused around the wavelength region of the \ion{Ca}{II} H \& K absorption lines, designed to efficiently search for extremely metal-poor stars. In this work, we use the first results of a medium-resolution spectroscopic follow-up to refine the selection criteria for finding extremely metal-poor stars ($\textrm{[Fe/H]} \leq -3.0$) in the \emph{Pristine} survey. We consider methods by which stars can be selected from available broad-band and infrared photometry plus the additional \emph{Pristine} narrow-band photometry. The spectroscopic sample presented in this paper consists of 205 stars in the magnitude range $14 < V < 18$. Applying the photometric selection criteria cuts the sample down to 149 stars, and from these we report a success rate of 70\% for finding stars with $\textrm{[Fe/H]} \leq -2.5$ and 22\% for finding stars with $\textrm{[Fe/H]} \leq -3.0$. These statistics compare favourably with other surveys that search for extremely metal-poor stars, namely an improvement by a factor of $\sim 4-5$ for recovering stars with $\textrm{[Fe/H]} \leq -3.0$. In addition, \emph{Pristine} covers a fainter magnitude range than its predecessors, and can thus probe deeper into the Galactic halo. 
\end{abstract}




\section{Introduction}

During Big Bang nucleosynthesis, only the lightest elements were produced in any significant quantity: mainly hydrogen, helium, and trace amounts of lithium. Nearly all heavier elements were formed later in the interiors of stars and released in their supernovae explosions, thereby enriching the metal content of the Universe over time \citep{Alpher48,Burbidge57}. The oldest stars that formed in the early Universe from this pristine gas should therefore be largely free from heavier elements due to their early time of formation. Analysis of these metal-poor stars, their stellar parameters, chemical abundances, dynamics, and spatial distributions in the Galaxy can offer insight into the local environments in which they formed, and thus help to invoke constraints on our understanding of the first generation stars that came before them and the Galaxy at early times \citep[e.g.,][]{Freeman02,Beers05,Frebel15}. 

Such stars are rare among the overwhelming numbers of more metal-rich populations in the Galaxy. From the Besan\c{c}on model of the Galaxy \citep{Robin03}, the expectation is that in a high Galactic latitude field towards the anti-centre direction ([l,b] = [0$^{\circ}$, 60$^{\circ}$]) and in the magnitude range $14 < V < 18$, only $\sim1/2000$ stars will be extremely metal-poor (EMP), with a metal content less than $1/1000$ of the Sun ($\textrm{[Fe/H]} \leq -3$). This ratio increases to $\sim1/500$ for $18 < V < 20$, as this fainter magnitude range probes more of the metal-poor halo rather than the metal-rich Galactic disk. Although these are only projections, since this model relies on assumptions about the EMP tail of the metallicity distribution function, they emphasize that an efficient pre-selection method is needed to find and study these very rare stars. This is one of the principal goals of the \emph{Pristine} survey. 

For stars close to the main sequence turnoff, broad-band optical colours hold some metallicity information due to line blanketing at blue wavelengths (\citealt{Schwarzschild55,Sandage_Eggen59,Wallerstein62,Ivezic08}). However, this relation typically breaks down at metallicities just below $\textrm{[Fe/H]} = -2$ \citep{starkenburg17a}, which is the metallicity range of greatest interest to study the oldest and most pristine star formation environments. More recent work has shown that with good $u-$band data, the SDSS photometric metallicities can be extended into the $\textrm{[Fe/H]} \sim -2.5$ regime when a technique of multiple fitting to calibrated isochrones is used \citep{An13,An15}. The Canada-France Imaging Survey (CFIS) has also shown increased success in photometric metallicity determination with their high quality $u-$band observations (Ibata et al., 2017, subm.). Although these recent advances have improved the capabilities of photometric metallicity calibrations, they still do not provide information for the EMP regime at $\textrm{[Fe/H]} \leq -3$. 

A recent study by \citet{Schlaufman_Casey} implemented a combination of optical and infrared broad-band filters from the Wide-Field Infrared Survey Explorer \citep[WISE,][]{WISE}, the Two Micron All-Sky Survey \citep[2MASS,][]{Skrutskie06}, and the AAVSO Photometric All-Sky Survey \citep[APASS,][]{Henden_2009,Henden_2014,Henden_2015} to photometrically identify EMP stars. Most of the selection power of this method is based on the strong molecular absorption down to $\textrm{[Fe/H]} = -2$ in the wavelength region covered by the WISE W2 filter (4.6 $\mu$), such that colour combinations with the WISE W1 (3.4 $\mu$) and the 2MASS J (1.2 $\mu$) filters can effectively select metal-poor stars from photometry alone (see Figure 1 from \cite{Schlaufman_Casey}). Nevertheless, due to the limiting magnitudes of the existing infrared photometry from WISE, and the quality cuts required for this method to work, this technique is mostly suited for very bright targets, and thus is mainly sensitive to local halo stars (both \citealt{Schlaufman_Casey} and \citealt{Casey15} adopt a faint threshold of $V=14$ for their sample). 

Whenever large samples of stars are targeted at a certain phase in their stellar evolution, their relative brightness will trace different distances and hence various environments inside our Galaxy. Many of the most metal-poor stars known have magnitudes brighter than $V=16$, due to the techniques with which they were discovered. Up until the last decade, the main sources for extremely metal-poor (EMP; $\textrm{[Fe/H]} \leq -3$) stars, as well as a few ultra metal-poor (UMP; $\textrm{[Fe/H]} \leq -4$) stars, were through Ca H \& K objective-prism surveys such as the HK survey \citep[with a magnitude limit of $B\sim15.5,$][]{Beers85,Beers92} and the Hamburg ESO survey \citep[HES, with a magnitude limit of $B \sim 17-17.5$,][]{Christlieb02}. In line with expectations, the fainter HES was more successful in finding EMP and UMP stars because it reached deeper into the metal-poor outer halo. 

Although magnitude ranges often limit the distance range probed, there are still significant differences in the chemical properties of EMP stars and their present-day location and kinematics in the Galaxy \citep{Cayrel_2004,Frebel06,Bonifacio_2009,Carollo10,Carollo12,An13,Starkenburg13,Skuladottir15}. This dependence of chemical composition on Galactic environment has been further emphasized by recent studies in the bulge of the Galaxy, namely that EMP stars in the halo are often enhanced in carbon, whereas EMP stars in the bulge rarely exhibit carbon enhancement \citep{Howes15, Howes16, Koch16, Lamb17}.

The need for large samples of EMP stars across various environments and magnitude ranges has been somewhat mitigated in recent years by large scale, blind spectroscopic surveys. Some examples of these include the Sloan Digital Sky Survey \citep[SDSS,][]{York_2000}, as well as its dedicated constituent spectroscopic campaigns, the Sloan Extension for Galactic Understanding and Exploration \citep[SEGUE,][]{Yanny_SEGUE,FernandezAlvar15,FernandezAlvar16}, and the Baryonic Oscillations Spectroscopic Survey \citep[BOSS,][]{Eisenstein_2011,Dawson_2013} \citep[for higher resolution follow-up of metal-poor stars based on these samples, see][]{Caffau13,Aoki13,AllendePrieto15a,Aguado16}. These large surveys have the advantage that they probe deeper than the objective-prism surveys, and can obtain large numbers of good quality spectra. Nevertheless, the success rates for discovering metal-poor stars in these surveys are naturally low because they do not specifically target these stars, and EMP stars compose a very small fraction of the total stellar content of the Galaxy. 

Several previous studies have described the approach of targeted narrow-band photometry on the \ion{Ca}{II} H \& K wavelength region as a means of providing metallicity information \citep[e.g.,][]{Twarog91,Twarog2000}. One recent example which has been particularly successful is the \emph{Skymapper} survey in the Southern Hemisphere \citep[e.g.,][]{Keller07}, which uses a $v$ filter \citep[wavelength coverage $\sim 3650-4000$ \AA,][]{Bessell2011} to photometrically pre-select metal-poor star candidates for spectroscopic follow-up. Operating on a similar concept, the \emph{Pristine} survey uses a specially designed filter (wavelength coverage $\sim 3900-4000$ \AA) that is even narrower and more targeted on the \ion{Ca}{II} H \& K absorption lines. Although it covers less sky area compared to \emph{Skymapper}, the \emph{Pristine} survey is better suited to efficiently study fainter targets because it utilizes the 4m-class Canada-France-Hawaii Telescope (CFHT), which provides a large aperture and excellent image quality, and is located in the Northern Hemisphere where SDSS broad-band photometry is readily available.  

In this work we use the first results of medium resolution follow-up spectroscopy of 205 stars within the \emph{Pristine} survey to assess the performance of the survey's photometric pre-selection. This sample consists of targets in the magnitude range $14 < V < 18$, and therefore represents only the brighter end of the full \emph{Pristine} target sample. We use this sample to assess and improve the criteria used for selecting follow-up candidates for spectroscopy. By doing so, we pave the way to the successful follow-up of even fainter targets, opening up the possibility of the \emph{Pristine} dataset to be used to efficiently select targets for large multiplexing spectroscopic surveys in the near future, such as the William Herschel Telescope Enhanced Area Velocity Explorer \citep[WEAVE,][]{Dalton12,Dalton14,Dalton_2016}, the 4-metre Multi-Object Spectroscopic Telescope \citep[4MOST,][]{deJong16}, the Subaru Prime Focus Spectrograph \citep[PFS,][]{Takada14}, or the Maunakea Spectroscopic Explorer \citep[MSE,][]{McConnachie16}. These survey efforts are expected to probe more of the pristine environments in the outskirts of the Galactic halo; one of the regions in the Galaxy expected to harbour possible first star environments, as highlighted in recent analyses of cosmological simulations \citep[e.g.,][]{Starkenburg17b}. 

The paper is organized as follows: In Sections \ref{sec:survey} and \ref{sec:followup} we introduce the \emph{Pristine} survey and its spectroscopic follow-up programme. In Section \ref{sec:improve}, we summarize and discuss improvements to our candidate selection criteria, as well as investigate whether infrared photometry and regularized regression techniques can be used to further improve our results. In Section \ref{sec:spectroscopic_results}, we present the results of the medium resolution spectroscopic follow-up, in particular comparing the predicted photometric metallicities from \emph{Pristine} to spectroscopically determined metallicities. Finally, in Section \ref{sec:discussion} we discuss the current purity and success rates of our target selection, compare them to expectations and other works, and discuss projections and strategies for the continuation of the \emph{Pristine} survey. In this paper, we demonstrate that the \emph{Pristine} survey shows unparalleled efficiency for finding the most metal-poor stellar populations of the Galaxy. This is key for the eventual completion of two of the survey's main objectives, which include finding large numbers of EMP stars to contribute to the characterization of the extremely metal-poor tail of the metallicity distribution function, as well as uncovering the exceedingly rare UMP stars.

\section{The \emph{Pristine} survey}\label{sec:survey}

For a full and detailed description of the \emph{Pristine} survey we refer the reader to \citet[][the first \emph{Pristine} survey paper, hereafter referred to as Paper I]{starkenburg17a}. Here, we recapitulate the essential elements of the survey.

\begin{figure*}
\includegraphics[width=\textwidth]{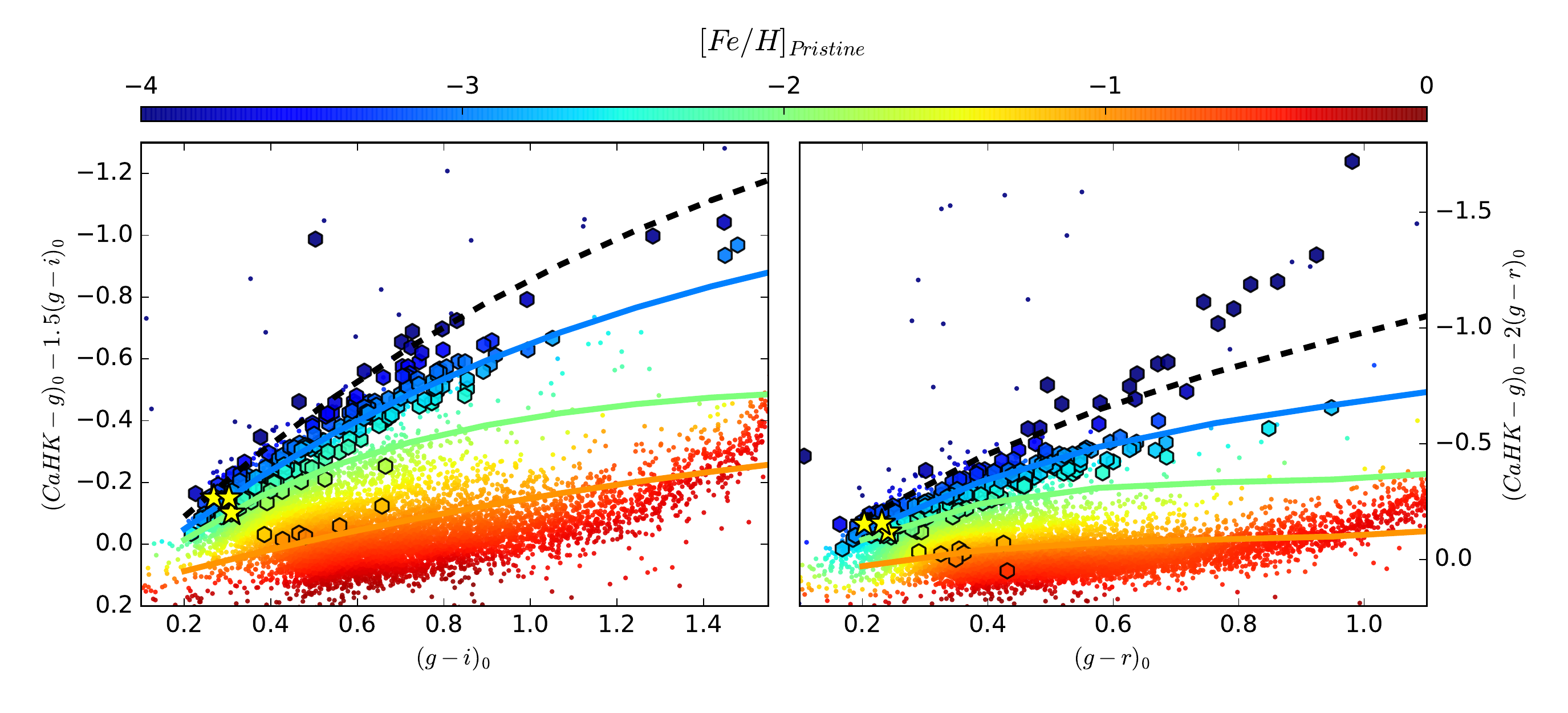}
\caption{The colour-colour space used to assign photometric metallicities for the \emph{Pristine} sample. The left and right panels show the calibrations for determining photometric metallicities using SDSS $g-i$ and $g-r$ colours, respectively. Stars that are included in the spectroscopic follow-up sample presented in this paper are shown with large symbols. The coloured lines trace constant metallicities of $\textrm{[Fe/H]} = -1$, $-2$, and $-3$, and the black dashed line represents the expected limit for stars that have no metal absorption lines in their spectra. The small points are 10,000 randomly selected \emph{Pristine} stars, to show the parameter space covered by the survey, and the yellow stars are the 3 stars for which example spectra are shown in Figure \ref{sample_spectra}. All symbols are coloured according to their derived photometric metallicities (see text for details).  \label{colour_colour_plot_Pristine}}
\end{figure*} 		

The \emph{Pristine} survey uses a narrow-band \ion{Ca}{II} H \& K filter (hereafter referred to as the $CaHK$ filter)  mounted on MegaPrime/MegaCam at the Canada France Hawaii Telescope (CFHT) on Mauna Kea in Hawaii. The filter was specifically designed by members of the \emph{Pristine} team to cover the wavelength region of the singly ionized \ion{Ca}{II} H \& K lines, located at 3968.5 and 3933.7 \AA, respectively. The narrow width of the filter reduces the influence of other spectral features, such as the nearby CN molecular absorption bands at 3839 and 4142 \AA. MegaPrime/MegaCam fields are $\sim 1 \textrm{ deg}^2$, and with integrations of 100 seconds, a signal to noise ($S/N$) of 10 at a depth of $g_0 \sim$ 21.0 can be achieved (Paper I). As of September 2016 the sky coverage was $\sim 1,000 \textrm{ deg}^2$, and data collection is ongoing with the aim to cover at least $\sim 3,000 \textrm{ deg}^2$. The footprint of the survey targets the Galactic halo and intentionally spans a range in Galactic latitude (30$^{\circ} < \mathrm{b} < 78^{\circ}$) to sample a diverse range of halo environments. Observations are made in the Northern Hemisphere, and overlap by design with regions of sky previously observed photometrically by SDSS. \emph{Pristine} can therefore cross-match its targets with SDSS to obtain $ugriz$ broad-band photometry, which is useful for temperature determination and point source identification, allowing for the elimination of most objects that are not stars. Another important advantage to the overlap with SDSS is that there is a sample of several thousand stars distributed over the \emph{Pristine} footprint for which moderate resolution (R $\approx$ 1800) spectra are already available from the SDSS and SEGUE surveys. Thus, \emph{Pristine} has a large sample of spectroscopic metallicities which can be used to calibrate the assignment of photometric metallicities, thereby greatly reducing the amount of overhead and telescope time required to make the survey operational. 

Figure \ref{colour_colour_plot_Pristine} depicts the parameter space used for assigning photometric metallicities. The y-axis shows the colour of the SDSS $g$-band minus the $CaHK$ magnitude obtained from the \emph{Pristine} narrow-band filter. An extra combination with the SDSS $g$ and $i$ or $r$ magnitudes is added to stretch the plot vertically and make it easier to see the metallicity gradient. The x-axis displays the SDSS $g-i$ or $g-r$ colours, which are proxies for stellar effective temperature. Unless specified otherwise, all magnitudes from SDSS and \emph{Pristine} discussed in the text of the rest of this paper refer to the deredened magnitudes (see Paper I for details on the deredening procedure). Lines of constant metallicity are also plotted, with orange, green, and blue representing [Fe/H] of $-1$, $-2$, and $-3$, respectively. These lines were produced using synthetic spectra models, generated with Model Atmospheres in Radiative and Convective Scheme \citep[MARCS,][]{Gustafsson_2008} stellar atmospheres and the Turbospectrum code \citep{Alvarez_Plez_1998,Plez_2008}. The black dashed line shows the theoretical limit for stars with no metal absorption lines present in their spectra (see Paper I for details). Using information from both these synthetic models and all stars with overlapping \emph{Pristine} photometry and SDSS/SEGUE spectra, this colour-colour space is divided into different photometric metallicity bins. \emph{Pristine} stars are then assigned a metallicity depending on the bin in which they fall, corresponding to their position in this plot. This procedure is followed for both $g-i$ and $g-r$ colours. The minimum metallicity that can be assigned is $-4.0$, and any object that falls outside of the calibrated regions (approximately the areas shown in Figure \ref{colour_colour_plot_Pristine}, up to 0.2 dex above the black dashed lines) receives a metallicity of -99. In cases where reliable metallicities are derived for both $g-i$ and $g-r$, the $g-i$ metallicity is preferentially used, since the sample space spans a larger colour range than in $g-r$, and therefore separates the sample more effectively by metallicity over the same range in temperature. We find that the photometric metallicity calibration has a standard deviation of 0.2 dex when compared with the spectroscopic metallicities from SDSS/SEGUE for the metallicity range from $\textrm{[Fe/H]} = -0.5$ down to $\textrm{[Fe/H]} = -3.0$ (Paper I). 

The small coloured points shown in Figure \ref{colour_colour_plot_Pristine} are a random selection of 10,000 \emph{Pristine} stars coloured according to their photometrically derived [Fe/H] values. Large hexagons are the 205 stars from the medium resolution follow-up sample used in this paper (see Section \ref{sec:followup}), also coloured by their corresponding \emph{Pristine} photometric metallicities. These stars are almost all selected from the upper regions of the plot between the $\textrm{[Fe/H]} = -2$ line, the $\textrm{[Fe/H]} = -3$ line, and the black dashed (no-metals) line, which are the regions expected to contain the most promising EMP star candidates. The stars that lie significantly above the no-metals lines -- particularly in the $g-r$ panel -- are either stars that were chosen before the full selection criteria described in this paper were implemented, or are stars that have moved in the plot as a result of improvements to the photometric reduction pipeline and calibration.

\section{Spectroscopic follow-up}\label{sec:followup}
In conjunction with the photometric component of \emph{Pristine}, a spectroscopic follow-up programme has been observing the most promising, bright ($V < 18$) metal-poor candidates on $2-4$m class telescopes with medium- and high-resolution spectrographs. In this paper, we focus on the homogeneous follow-up sample of 205 candidate stars observed with the Intermediate Dispersion Spectrograph (IDS) on the 2.5m Isaac Newton Telescope (INT) over the period of March 18 - 27, May 15 - 23, July 20 - 24, and September 2 - 6, 2016, and with the Intermediate dispersion Spectrograph and Imaging System (ISIS) on the 4.2m William Herschel Telescope (WHT) over the period of May 1-2, and July 29-31, 2016 (Programs C71 and N5). Both telescopes are located at the Roque de Los Muchachos Observatory in La Palma, Canary Islands. For the INT, the EEV10 CCD and the R900V grating with a 1.0" slit width were used, resulting in a resolution of 3333 at 4500 \AA\ over 2 pixels at the detector. For the WHT, the R600b and R600R gratings were used, along with the GG495 filter in the red arm. In conjunction with the default dichroic (5300) and a 1.0" slit, the set-up provided a mean resolution of 2400 and 5200 in the blue and red arms, respectively.

\begin{figure*}
	\includegraphics[width=\textwidth]{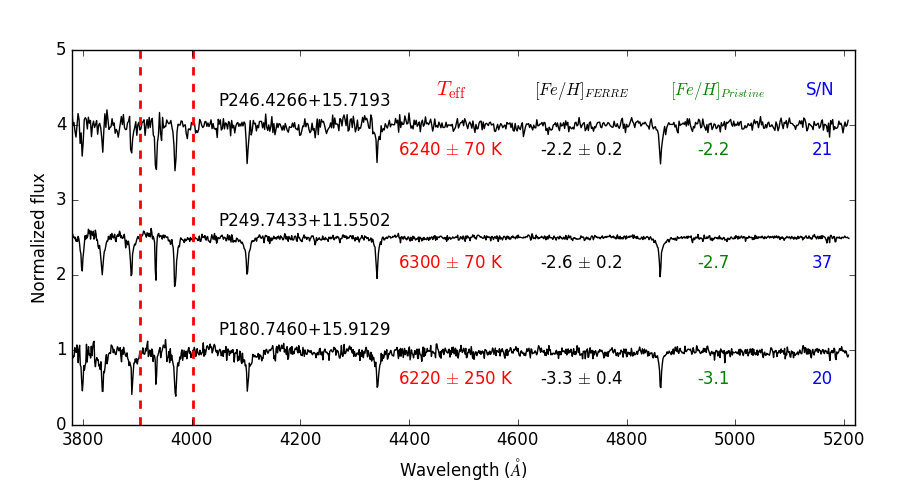}
    \caption{Sample spectra of three \emph{Pristine} target stars with different metallicities but similar temperatures, as determined spectroscopically by FERRE. These stars are marked in the \emph{Pristine} colour-colour space as yellow stars in Figure \ref{colour_colour_plot_Pristine}. The dotted red lines show the wavelength region (limits at which the transmission falls below 50\% of the maximum) for the $CaHK$ filter, and demonstrate the sensitivity of the filter to detecting changes in the strength of the \ion{Ca}{II} H \& K lines.}
    \label{sample_spectra}
\end{figure*}

\subsection{Data reduction and analysis}

Spectra were reduced using the Image Reduction and Analysis Facility  \citep[IRAF,][]{Tody_IRAF} software package. All basic reduction steps were implemented, including: image preprocessing (bias subtraction, flat fielding), spectrum extraction, sky subtraction, wavelength calibration, and heliocentric radial velocity correction. Although fringing has been shown to sometimes be a problem for the EEV10 CCD on the INT, the amplitude is lower than 5 \% when $\lambda < 6500$ \AA\footnote{\url{http://www.ing.iac.es/astronomy/instruments/ids/ids_eev10.html}}. The spectral range used for stars in our sample cover a wavelength range of $\sim 3750 - 5210$ \AA, therefore, it was not necessary to apply a fringing correction.

\subsection{Spectral analysis}

The spectra were analyzed using FERRE\footnote{FERRE is available from \url{http://github.com/callendeprieto/ferre}} 
\citep{Allende_Prieto_2006}. We provide here some basic information about the analysis process, but for a detailed account we refer the reader to \citep[]{Aguado17a,Aguado17b}. 
Both the observed and synthetic spectra were normalized using a running-mean filter 30 pixels wide. The FERRE code fits the entire available spectral region and searches for the atmospheric parameters that best match the observed spectrum by interpolating within the grid. The grid of synthetic spectra used is similar to the one described by \citet{Allende_Prieto_2014}, but with [C/Fe] as a free parameter \citep{AllendePrieto15b,Aguado16,Aguado17a,Aguado17b}. This grid has four dimensions and limits of $-6 < \textrm{[Fe/H]} < -2$, $1 < \textrm{log }g < 5$, $-1 < \textrm{[C/Fe]} < 5$, and $4750\rm K < T_{\rm eff} < 7000\rm K$.

To determine the uncertainties, the metallicities are re-derived 50 times after injecting random noise, according to the noise model provided by the data reduction pipeline, and a normal distribution for each instance. The standard deviation of the resulting metallicity distribution is taken as the metallicity uncertainty. Following \citet{Aguado17b} we add an additional 0.1 dex to the uncertainties to account for other systematic effects.

Figure \ref{sample_spectra} illustrates some typical spectra obtained from the INT, and the relevant wavelength region used for the analysis, $\sim 3750 - 5210$ \AA. The three sample spectra shown were specifically chosen to have similar temperatures, such that the line absorption in the wavelength region targeted by the narrow-band filter can easily be compared. Both the \ion{Ca}{II} H (3968.5 \AA) and the \ion{Ca}{II} K (3933.7 \AA) lines are weaker in more metal-poor stars of similar stellar parameters. In relatively warm stars, such as those shown here, the \ion{Ca}{II} H line remains somewhat stronger, since it is blended with the H$_{\epsilon}$ line (3970 \AA). Therefore, it is particularly the \ion{Ca}{II} K line that is a good indicator of whether a star is deficient in all metals, including calcium \citep[e.g.,][]{Beers99}. We note that at this resolution, we cannot typically resolve the interstellar calcium lines form the \ion{Ca}{II} H \& K lines. However, since any additional blended features only increase the strength of the lines, this will result in stars appearing more metal rich than they actually are, but not more metal-poor.

\section{Selection criteria}
\label{sec:improve}

One of the main goals of this paper is to assess and improve the selection of spectroscopic follow-up stars based on the photometric parameters. Throughout the spectroscopic follow-up, we have developed a specific set of criteria to remove the most contaminants while keeping the completeness as high as possible. These criteria are described in this section. 

\subsection{SDSS photometry}
\label{sec:SDSS_photometry}

SDSS was chosen as the principal survey to combine to \emph{Pristine} because of its large footprint in the Northern Hemisphere and excellent quality of well-calibrated, deep broad-band photometry. We evaluate the photometric information in several of the SDSS broad-band filters combined with the \emph{Pristine} narrow-band information. The selection criteria that we refined with the spectroscopic sample are described below:

\begin{itemize}
	\item{\textit{Non-star contamination}: Objects that are not stars may exhibit strange spectral signatures that could make them appear to be metal-poor stars from our photometric selection. We therefore identify and remove as many of these sources as possible during the photometric reduction to minimize this source of contamination. The photometry was reduced using the Cambridge Astronomical Survey Unit pipeline \cite[CASU,][]{Irwin_2001}, and modified to work specifically for CFHT/MegaCam data \citep{Ibata_2014}. Objects identified as being stars are flagged as such, and we impose that requirement for objects to be considered for further follow-up. In addition, when matching \emph{Pristine} to SDSS, we only consider sources labelled as stars, thereby providing another means to remove non-point source objects.}
	\item{\textit{White dwarf contamination}: Most white dwarfs have very weak CaHK absorption features and therefore could be mistaken for metal-poor stars with the \emph{Pristine} narrow-band filter. Stars at $u-g$ magnitude $< 0.6$ are likely to be white dwarfs, and are as such easily separated from most main-sequence and giant stars \citep[Ibata \emph{et al.} 2017, subm.]{Lokhorst}. We use this colour cut to remove white dwarfs from the sample.}
	\item{\textit{Variability}: Since the SDSS $ugriz$ broad-band observations and the \emph{Pristine} narrow-band $CaHK$ observations were taken several years apart, any variable objects could show large variations in brightness between the two data acquisitions, and therefore move significantly in the vertical direction on the colour-colour plot shown in Figure \ref{colour_colour_plot_Pristine}. This would result in the scattering of non-metal-poor stars into the metal-poor regime and contaminate the sample of stars selected for follow-up. In order to remove these variable objects, the Chi-square variability parameter measured from Pan-STARRS1 photometry was used \citep{Paanstars_variable}, namely that the Pan-STARRS1 variability flag $< 0.5$. It should be noted that this variability index is only sensitive to brightness variations over the period of the Pan-STARRS1 survey. Since this timescale is shorter than the difference in time between the SDSS and \emph{Pristine} observations, this flag will fail to remove variable objects with periods longer than the Pan-STARRS1 survey. Thus, these objects remain as a source of contamination, although the total number of these in our sample is expected to be quite small.}
	\item{\textit{Quality of SDSS i-, r-, and g-band photometry}: We consider SDSS photometric quality flags for saturation, blending, interpolation problems, objects too close to the edge of the frame, or suspicious detections, in each of the $g$-, $r$- and $i$-bands. Since our sample is crossmatched with SDSS and metallicity determinations depend upon the $g-i$ and $g-r$ colours, stars flagged in SDSS as having bad photometry that affects both of these colours must be removed from the sample. We therefore immediately remove all stars which are flagged for bad photometry in the $g$-band. For the $r$- and $i$-bands, we only remove a star if it is flagged as having bad photometry in both bands, because if only one of them is flagged, we may still be able to obtain a reliable photometric metallicity from the other.}
		\item{\textit{Probability of a star to have $\textrm{[Fe/H]} \leq -2.5$ in both $g-r$ and $g-i$}: We choose specifically not to use computed uncertainties in the metallicities because a Monte Carlo estimated uncertainty probability distribution in metallicity space is distinctly non-Gaussian in shape. Instead, we compute probabilities to reflect the likelihood that a given star has an $\textrm{[Fe/H]}_{Pristine} \leq -2.5$. To compute these, we take the uncertainties in $CaHK$, $g$, and $i$ or $r$ photometry, and re-draw these magnitudes in a Monte Carlo fashion for $10^4$ instances. For each re-draw, the \emph{Pristine} photometric metallicity is calculated from the fiducial $CaHK$, $g$, and $i$ or $r$ magnitudes. The probability of a star to have $\textrm{[Fe/H]}_{Pristine} \leq -2.5$ is subsequently determined by the fraction of the draws for which it gets assigned a photometric metallicity below $\textrm{[Fe/H]}_{Pristine} = -2.5$. This  procedure is done for both the $g-i$ and $g-r$ photometric metallicities. We discard any star for which this probability is less than 0.25 for both $g-i$ and $g-r$.}
	\item{\textit{Photometric metallicity grid}: If a star falls outside of the parameter space for which the assignment of photometric metallicities has a valid calibration, it is assigned a metallicity of $-99$. This is approximately the region contained within Figure \ref{colour_colour_plot_Pristine}, up to 0.2 dex above the black dashed no-metals line. If a star has a metallicity of $-99$ for both $g-r$ and $g-i$, it is not considered for follow-up.  Similarly, if a star is assigned a metallicity of $-99$ for only one of $g-r$ or $g-i$, it is removed from the sample if it also has a probability $(\mathrm{[Fe/H] \leq -2.5}) < 0.25$, or bad photometry in the other band.}
	\item{\textit{Colour range}: The colour ranges over which \emph{Pristine} most successfully separates stars of different metallicities are $0.25 < g-i < 1.5$ and $0.15 < g-r < 1.2$. These colour ranges correspond roughly to temperatures of $4200\rm{K} < T_{\rm eff} < 6500\rm{K}$, covering the tip of the red giant branch and the cooler main sequence, all the way to the main sequence turn-off. For hotter stars, the different [Fe/H] populations exhibit more overlap and thus assignment of a metallicity in this regime suffers from larger uncertainty and is more susceptible to contamination by more metal-rich stars. For cooler stars, the main-sequence population at the $\textrm{[Fe/H]} = -1$ line begins to turn upward and contaminate the more metal-poor red giant regimes. Some stars that fall outside of these colour ranges may still be assigned valid photometric metallicities and may still be interesting targets, but these are followed-up at a lower priority because these regions have a higher contamination rate.}
\end{itemize}

To summarize the selection criteria, a star is removed from the sample if any of the following are true:
\begin{itemize}
	\item{P($\mathrm{[Fe/H]}_{g-r} \leq -2.5) < 0.25$ and P($\mathrm{[Fe/H]}_{g-i} \leq -2.5) < 0.25$ (1)}
	\item{$g$-band phot flag (2)}
	\item{point source flag (CASU flag) $\neq -1$ (3)}
	\item{$r$-band phot flag  and P($\mathrm{[Fe/H]_{g-i} \leq -2.5}) < 0.25$ (4)}
	\item{$i$-band phot flag  and P($\mathrm{[Fe/H]_{g-r} \leq -2.5}) < 0.25$ (5)}
	\item{$\mathrm{[Fe/H]}_{g-i}  = -99$ and P($\mathrm{[Fe/H]}_{g-r} \leq -2.5) < 0.25$ (6)}
	\item{$\mathrm{[Fe/H]}_{g-r} = -99$ and P($\mathrm{[Fe/H]}_{g-i} \leq -2.5) < 0.25$ (7)}
	\item{$\mathrm{[Fe/H]}_{g-r} = -99$ and $\mathrm{[Fe/H]}_{g-i} = -99$ (8)}
	\item{$i$-band phot flag and $r$-band phot flag (9)}
	\item{$u-g$ mag $< 0.6$ (10)}
	\item{variability $> 0.5$ (11)}
	\label{quality_cuts_list}
\end{itemize}

\addtocounter{footnote}{-1}
\begin{figure}
	\includegraphics[width=\columnwidth]{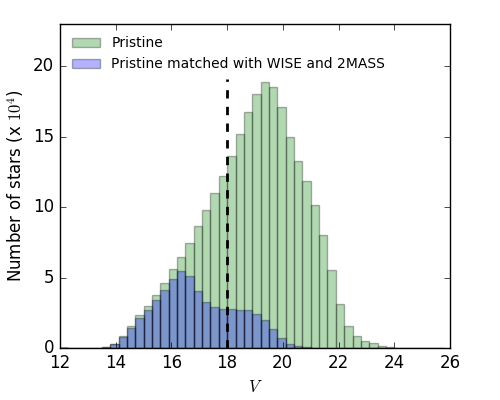}
        \caption[]{Distribution of $V$ magnitudes\footnotemark{} for \emph{Pristine} stars (green), and stars that have available WISE and 2MASS magnitudes and satisfy the quality cuts used in SC14 (blue). The black dashed line shows the magnitude limit for the spectroscopic sample in this paper.}
    \label{wise_match_hist}
\end{figure}
\footnotetext{$V$ magnitudes are calculated using SDSS g and r magnitudes according to the relation described in \url{https://www.sdss3.org/dr8/algorithms/sdssUBVRITransform.php} (Lupton 2005)}

\subsection{Infrared magnitudes from WISE and 2MASS}
A study conducted by \citet[hereafter referred to as SC14]{Schlaufman_Casey} has shown that there is metallicity information contained in the infrared wavelength regions. In their paper, they devise a set of novel selection criteria using the infrared broad-band filters of WISE and 2MASS to select for metal-poor stars. The main selection power of this method comes from the WISE W2 band, centred at 4.6 $\mu$, which contains molecular bands that are strongly metallicity dependent. In this section, we investigate whether the addition of this infrared magnitude information could increase the selection efficiency of \emph{Pristine}. 

We applied the selection criteria from SC14 to the \emph{Pristine} photometric sample with $V < 18$, and the analysis revealed two significant limitations on its ability to improve the \emph{Pristine} selection. The first of these is that the WISE and 2MASS magnitudes are only available for a subset of the brightest stars in the \emph{Pristine} sample. Figure \ref{wise_match_hist} illustrates the overlap between the total \emph{Pristine} SDSS-matched sample and the stars for which WISE and 2MASS broad-band information is available. To this sample, we have also applied the quality cuts defined in SC14 (only the flag criteria, but not the colour criteria; see their appendix), such that it is a true representation of the subsample of \emph{Pristine} for which this analysis could be performed. When applied to the brightest subset of this sample ($V < 15$), the WISE and 2MASS selection criteria increased the relative number of stars with [Fe/H] < -2.5 from 0.7\% to 3.9\%, where the metallicities are those derived photometrically from \emph{Pristine}. In the $15 < V < 16$ magnitude bin, the improvement was less pronounced (from 0.7\% to 1.7\%), and for the fainter magnitude samples ($V > 16$) this selection power was completely lost. This is to be expected, as the uncertainties in the WISE catalogue for these fainter magnitudes quickly become larger than the range allowed by the selection criteria (e.g., $-0.04 \leq W1-W2 \leq 0.04$). To account for this, SC14 limit their sample to bright stars with $V < 14$. SDSS photometry is limited to $V \gtrsim 14$ (for a typical star this corresponds roughly to a $CaHK \sim 15$) due to saturation, and because \emph{Pristine} is matched with SDSS, it also inherits this limit. In principle, \emph{Pristine} can observe brighter stars, down to a magnitude of $CaHK \sim 12$, but for these stars the narrow-band information must be used in conjunction with broad-band photometry with a brighter saturation limit than SDSS. This has been successfully demonstrated using APASS, in a recent paper by the \emph{Pristine} collaboration \citep{Caffau17}. However, even for the bright samples where SC14 provides selection power, it removes a large number of the stars of primary interest with $\textrm{[Fe/H]} \leq -2.5$. Therein lies the second major limitation of the SC14 selection criteria and its application to the \emph{Pristine} sample, its low completeness in the metal-poor regime.

\begin{figure*}
	\includegraphics[width=\textwidth]{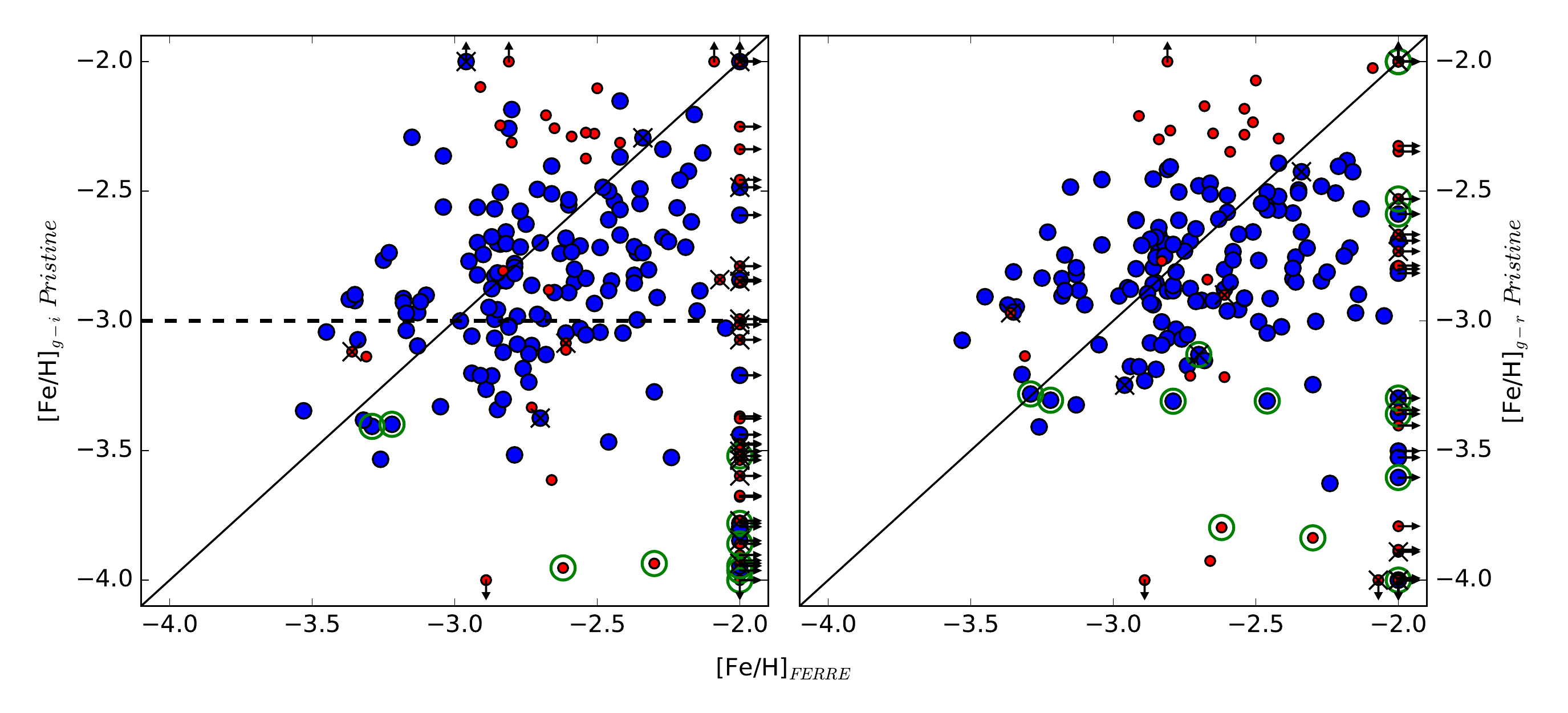}
    \caption{Photometric metallicities for both $g-i$ and $g-r$ plotted against spectroscopic metallicity. The data points are labelled as follows: stars that pass the selection criteria (blue large circles), stars that do not pass the selection (red smaller circles), stars with bad photometry in $i$ for the left panel and $r$ for the right panel (marked with an x), and stars that are above the theoretical no-metals line in Figure \ref{colour_colour_plot_Pristine} (circled in green). Data points that fall outside of the plotted region are forced to the border of the plot and marked with arrows showing their true positions.}
    \label{pristinevsFERRE}
\end{figure*}

For these reasons, we conclude that the WISE and 2MASS selection criteria as implemented by SC14 are quite limited in their application to the \emph{Pristine} sample, and we therefore do not include them in our selection criteria. However, if a large sample of bright, moderately metal-poor candidates are indifferentiable by the \emph{Pristine} selection criteria and follow-up telescope time is limited such that they can not all be observed, then the WISE and 2MASS selection criteria may be useful as a final means to prioritize the sample.

\begin{figure*}
	\includegraphics[width=\textwidth]{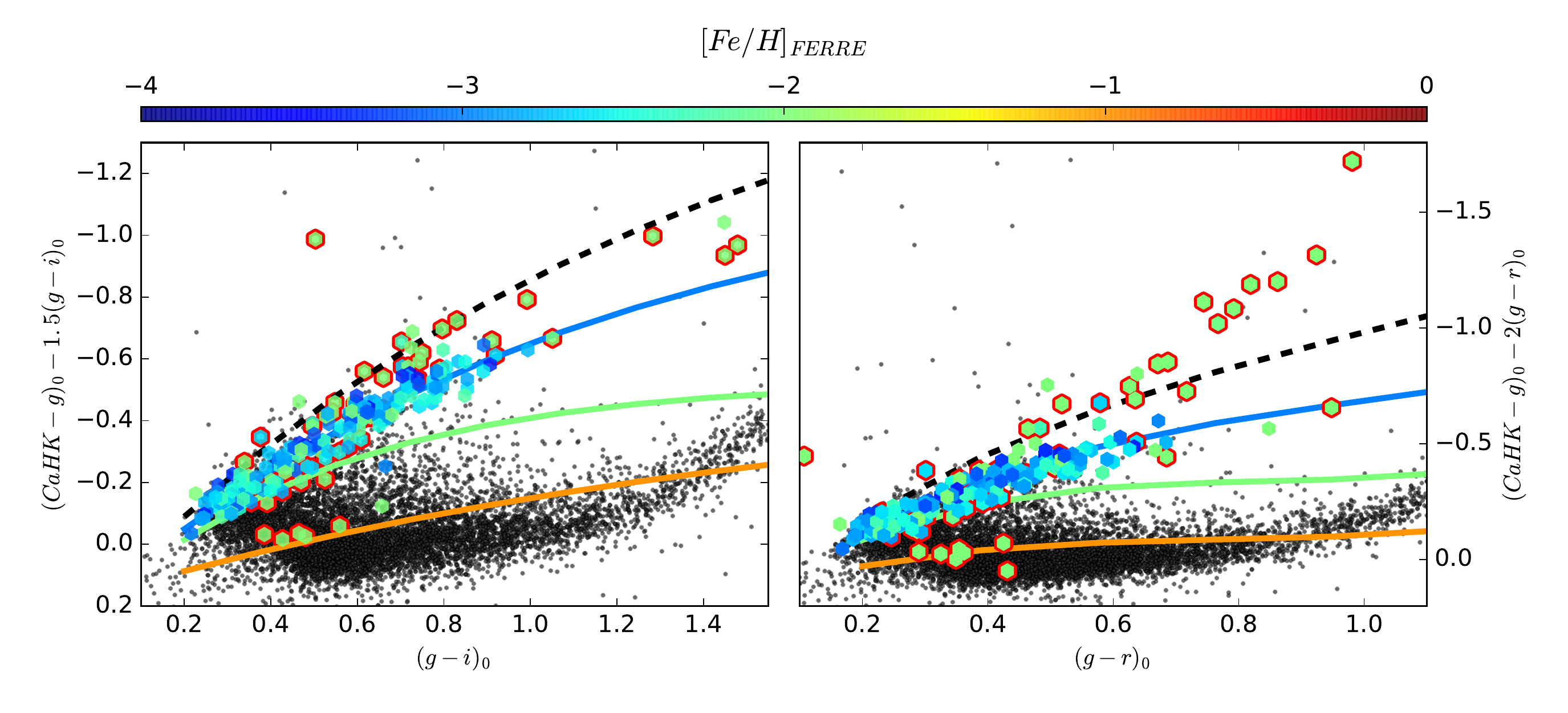}
	\caption{The same colour-colour space as in Figure \ref{colour_colour_plot_Pristine}, but with stars that have been selected for spectroscopic follow-up coloured according to their spectroscopic $\textrm{[Fe/H]}_{\textrm{FERRE}}$. The coloured lines trace along constant metallicities and the black dashed line is the expected limit of stars that have no metal absorption lines in their spectra. The grey points are 10,000 randomly selected \emph{Pristine} stars, to show the parameter space covered by the survey. Data points circled in red represent the stars that are removed from the sample by the selection criteria, many of which are contaminants with $\textrm{[Fe/H]}_{\textrm{FERRE}} \geq -2$. These stars circled in red are listed at the bottom of Table \ref{cut_criteria_table}.} 
	\label{colour_FERRE}
\end{figure*}

\subsection{Applying a regularized regression technique}
We applied a regularized regression technique, namely Lasso LARS \citep{tibshirani1996,efron2004}, to further assess the need to add other photometric data to predict [Fe/H] metallicities. Such a technique can tell us the leverage of any photometric colour or flux, thereby giving us an independent -- and unbiased -- view on the most valuable photometric information. It defines a model from a polynomial combination of all the photometric inputs, which also includes colours and cross-terms between the different bands. While doing so, the regularization in this method additionally acts to prefer solutions with fewer parameter values. Such a complete model allows us to effectively explore the importance of various datasets, such as WISE, 2MASS, Spitzer, and SDSS.

During this procedure, we found the $u-g$ colour to be efficient in flagging peculiar objects that could contaminate our sample, which corresponds to our usage of this colour to select out white dwarf contaminants. We also confirmed that adding infrared data such as Spitzer or WISE photometry does not contain significantly independent information from our initial SDSS $ugriz$ + $CaHK$ dataset. Indeed, the metallicity information seems to be mostly contained in the $CaHK - g$, and $g-i$ colour combinations, with extra information in the SDSS $u-$ and $r-$bands. 

At this stage of the Pristine survey, our training set (i.e., the cleaned photometric sample and the SDSS/SEGUE metallicities) does not contain many stars with $\textrm{[Fe/H]} \leq -2.5$. Therefore, it may be fruitful to repeat this analysis once a larger training set becomes available after additional follow-up spectroscopy. 

Taken together, the analysis of the infrared information contained in WISE and 2MASS and an analysis using regularized regression did not result in any changes to the selection criteria described at the end of Section \ref{sec:SDSS_photometry}. We therefore proceed with this list of selection criteria to choose stars for future follow-up spectroscopy.

\section{Spectroscopic results}
\label{sec:spectroscopic_results}

\begin{table*}
\caption{Metallicities of Pristine stars from photometry and spectroscopy. Column $CaHK$ is the magnitude obtained from the Pristine narrow-band filter, columns $\textrm{[Fe/H]}_{g-i}$ and $\textrm{[Fe/H]}_{g-r}$ are the photometric metallicities determined using the $g-i$ and $g-r$ colours, respectively. Each is followed by the corresponding derived probability that this metallicity is $\leq -2.5$. The next two columns are the spectroscopic metallicities derived form FERRE and their associated uncertainties. Column S/N is the signal to noise ratio of the analyzed spectrum, and column Inst. indicates the instrument used for the observations: either INT/IDS or WHT/ISIS. The last column shows which selection criteria a given star did not pass, and the flags are encoded according to the numbers assigned in the summary list of selection criteria in Section \ref{sec:improve}. We include here only 10 of the 205 stars observed, to show the form of the table. The full table is available online, along with a supplementary table which gives the SDSS coordinates and magnitudes for each of the stars in the sample.}
\centering
\begin{tabular}{cccccccccccc}
\hline 
\multicolumn{1}{|c|}{\textbf{Name}} & \multicolumn{1}{c|}{\textbf{$CaHK$}} & \multicolumn{1}{c|}{\textbf{($\pm$)}} & \multicolumn{1}{c|}{\textbf{[Fe/H]}} & \multicolumn{1}{c|}{\textbf{prob}} & \multicolumn{1}{c|}{[\textbf{Fe/H]}} & \multicolumn{1}{c|}{\textbf{prob}} & \multicolumn{1}{c|}{\textbf{[Fe/H]}} & \multicolumn{1}{c|}{\textbf{($\pm$)}} & \multicolumn{1}{c|}{\textbf{S/N}} & \multicolumn{1}{c|}{\textbf{Inst.}} &  \multicolumn{1}{c|}{\textbf{Flags}} \\
\multicolumn{1}{|c|}{} & \multicolumn{1}{c|}{} & \multicolumn{1}{c|}{} & \multicolumn{1}{c|}{g-i} & \multicolumn{1}{l|}{\textbf{[Fe/H]}} & \multicolumn{1}{c|}{g-r} & \multicolumn{1}{l|}{\textbf{[Fe/H]}} & \multicolumn{1}{c|}{\textbf{FERRE}} & \multicolumn{1}{c|}{} & \multicolumn{1}{c|}{} & \multicolumn{1}{c|}{}  & \multicolumn{1}{c|}{} \\
\multicolumn{1}{|c|}{} & \multicolumn{1}{c|}{} & \multicolumn{1}{c|}{} & \multicolumn{1}{c|}{} & \multicolumn{1}{c|}{g-i} & \multicolumn{1}{c|}{} & \multicolumn{1}{c|}{g-r} & \multicolumn{1}{c|}{} & \multicolumn{1}{c|}{} & \multicolumn{1}{c|}{} & \multicolumn{1}{c|}{} & \multicolumn{1}{c|}{} \\
\multicolumn{1}{|c|}{} & \multicolumn{1}{c|}{} & \multicolumn{1}{c|}{} & \multicolumn{1}{c|}{} & \multicolumn{1}{l|}{\textbf{$< -2.5$}} & \multicolumn{1}{c|}{} & \multicolumn{1}{l|}{\textbf{$< -2.5$}} & \multicolumn{1}{c|}{} & \multicolumn{1}{c|}{} & \multicolumn{1}{c|}{} & \multicolumn{1}{c|}{} & \multicolumn{1}{c|}{}  \\
\hline  
Pristine\_183.5424+13.6790 & 15.42 & 0.02 & -3.0 & 0.98  & -3.0  & 0.99  & $\geq-2.0$ & 0.2 & 24 & IDS  & -         \\
Pristine\_184.7471+10.6008 & 15.97 & 0.02 & -3.4 & 1.00  & -3.5  & 1.00  & $\geq-2.0$ & -   & 27 & IDS  & -         \\
Pristine\_185.0736+15.1006 & 15.84 & 0.02 & -2.8 & 0.97  & -2.8  & 0.97  & -2.4       & 0.2 & 27 & IDS  & -         \\
Pristine\_185.6263+06.1900 & 15.46 & 0.02 & -3.3 & 1.00  & -3.2  & 1.00  & -2.9       & 0.2 & 31 & IDS  & -         \\
Pristine\_186.5993+15.0468 & 15.49 & 0.02 & -2.5 & 0.57  & -2.6  & 0.63  & -2.4       & 0.3 & 27 & IDS  & -         \\
\hline
Pristine\_245.1095+08.8947 & 14.94 & 0.02 & -2.8 & 0.96  & -99   & -0.01 & -2.1       & 0.2 & 15 & IDS  & 2,5       \\
Pristine\_237.5278+12.2989 & 16.18 & 0.02 & -3.0 & 1.00  & -0.0  & 0.00  & $\geq-2.0$ & -   & 10 & IDS  & 3,5       \\
Pristine\_240.8957+08.4476 & 16.93 & 0.02 & -3.5 & 1.00  & -99   & -0.01 & $\geq-2.0$ & -   & 32 & ISIS & 3,5       \\
Pristine\_182.5908+06.1748 & 17.28 & 0.02 & -99  & -0.01 & -99   & -0.01 & -2.9       & 0.3 & 16 & IDS  & 1,3,6,7,8  \\
Pristine\_230.9962+07.4789 & 15.66 & 0.02 & -99  & -0.01 & -99   & -0.01 & $\geq-2.0$ & -   & 43 & IDS  & 1,3,6,7,8 \\
\hline
\hline
\end{tabular}
\label{cut_criteria_table}
\end{table*}

Figure \ref{pristinevsFERRE} shows the photometrically predicted [Fe/H] with \emph{Pristine} for both $g-i$ and $g-r$ and the spectroscopically determined [Fe/H] with FERRE for all of the stars followed up at the INT and WHT. In the following discussion of the results, we use the terms \emph{Pristine} metallicities and photometric metallicities synonymously to refer to the metallicity values derived from the narrow-band photometric \emph{Pristine} + SDSS $ugriz$ data, and the terms FERRE metallicities and spectroscopic metallicities to refer to the metallicities derived from analysis of the spectra with FERRE. Only spectra of sufficient quality to be reliably analyzed with FERRE (this was decided visually by the authors, but approximately follows a cut of $S/N = 10$) are included in the sample, which totals 205 stars. The blue large circles represent the 149 stars which pass all of the selection criteria summarized in the list at the end of Section \ref{sec:SDSS_photometry}, and the red smaller circles are removed on the basis of at least one of the selection criteria. In both panels, the \emph{Pristine} metallicities are skewed toward the metal-poor end when compared to FERRE. This reflects two characteristics of the sample: 1) stars predicted by photometry to be more metal-poor were preferentially selected for spectroscopic follow-up and 2) because of the shape of the metallicity distribution function, there will be more stars at higher metallicities that will scatter into our photometrically selected sample than the other way around. As a check, we looked to see if there was a correlation between $\textrm{[Fe/H]}_{Pristine}$ and the computed probabilities of having $\textrm{[Fe/H]}_{Pristine}\leq-2.5$. Indeed, these showed a tight anti-correlation, which was the expected behaviour, given that the current sample is relatively bright and has small photometric uncertainties.

All objects in Figure \ref{pristinevsFERRE} at $\textrm{[Fe/H]}_{\textrm{FERRE}} = -2.0$ should be interpreted as having $\textrm{[Fe/H]} \geq -2.0$, since the spectral grid used for this analysis was specifically optimized for metal-poor stars and only assigned metallicities in the range $-6 \leq \textrm{[Fe/H]} \leq -2$. Future work will extend the grid of synthetic spectra to higher metallicities to determine metallicity values for these more metal-rich stars, but for the purposes of this work it is sufficient just to classify them as contaminants. Points marked with an X are flagged with bad photometry in either the i- or r- bands, and points circled in green fall above the no-metals line in Figure \ref{colour_colour_plot_Pristine}. In both the $g-i$ and $g-r$ panels, most of the stars that fall above the no-metals line are contaminants located at $\textrm{[Fe/H]}_{\textrm{FERRE}} \geq -2$. However, a few of the stars that do remain are some of the most metal-poor in the sample, and therefore removing stars based solely on this criteria may be detrimental as it could potentially remove the very rare UMP stars that we are searching for. Dealing with the stars above this line is therefore a matter of completeness versus purity, and given that finding a large number of EMP stars and finding the extremely rare UMP stars are both major objectives of this survey, a choice needs to be made. With the current, small sample of stars that fall in this regime, it is difficult to make a quantitatively driven decision about this matter. Fortunately, many of the stars that fall above the no-metals line are already removed by other selection criteria. We therefore decide not to eliminate the stars that fall above the no-metals line from the sample, in order to mitigate the risk of missing potential UMP stars, but at the cost of a slightly increased contamination rate.

Figure \ref{colour_FERRE} shows the same colour-colour space as Figure \ref{colour_colour_plot_Pristine}, but with the spectroscopic sample coloured by their FERRE metallicities, with stars that do not meet the selection criteria highlighted in red. Again, stars with a light green colour corresponding to $\textrm{[Fe/H]}_{\textrm{FERRE}} = -2.0$ actually have $\textrm{[Fe/H]} \geq -2.0$ and are contaminants. Many of these are successfully removed with the implementation of the selection criteria. Finally, Table \ref{cut_criteria_table} tabulates the photometric and spectroscopic metallicities for all of the stars in the sample. The stars that do not pass all the selection criteria are listed last, with the rightmost column showing exactly which selection criteria they failed to meet. This table also provides the uncertainties for the spectroscopic metallicities. In this paper, we report and use only the [Fe/H] values of this sample in order to assess the follow-up success of Pristine. The full sample, as well as determinations of stellar parameters and other abundances are presented in a companion paper (Aguado et al., in prep.). 

It is clear from Table \ref{cut_criteria_table}, Figure \ref{pristinevsFERRE}, and Figure \ref{colour_FERRE} that there is still some useful information in the \emph{Pristine} photometry, even at these low metallicities of $\textrm{[Fe/H]} \leq -2.5$. In Figure \ref{pristinevsFERRE}, although there is a scatter around the one-to-one line, the lowest metallicity stars from their spectroscopic metallicities also typically have a lower photometric metallicity determination.

\section{Discussion}\label{sec:discussion}
\subsection{Purity and success rates of the selection}\label{sec:purityandsuccess}

The current spectroscopic sample can be divided into three groups: a total sample of all 205 stars that were observed, a subsample of the 149 stars that pass all of the selection criteria, and a subsample of the 46 stars with $\textrm{[Fe/H]}_{Pristine} \leq -3.0$, which represents the best candidates (all stars below the black-dotted line in the left panel of Figure \ref{pristinevsFERRE}). Table \ref{success_rates} presents the numbers of stars in various photometric and spectroscopic metallicity bins for these three samples. Firstly, it shows the number of stars in each sample that were predicted by \emph{Pristine} photometric metallicities to have $\textrm{[Fe/H]} \leq -2.5$ or $\leq -3.0$, respectively. In addition, it provides the same numbers according to the spectroscopic $\textrm{[Fe/H]}_{\textrm{FERRE}}$, and the success rates, which we define as the fraction of stars predicted to be below a certain $\textrm{[Fe/H]}_{Pristine}$ that were actually found to have $\textrm{[Fe/H]}_{\textrm{FERRE}}$ below that value. The selection criteria (as defined in Section \ref{sec:improve}) increase the relative fraction of metal-poor stars in all cases, and eliminate a large number of contaminants with $\textrm{[Fe/H]}_{\textrm{FERRE}} \geq -2$, as compared to the total sample. The $\textrm{[Fe/H]}_{Pristine} \leq -3.0$ sample increases the the relative fractions even more, but concedes a higher contamination rate than the sample which passes the selection criteria.

\begin{figure}
	\includegraphics[width=\columnwidth]{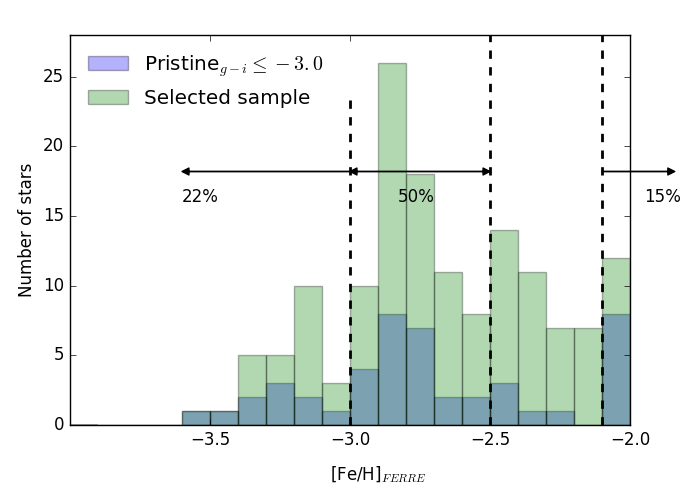}
    \caption{The FERRE metallicity distribution for the selected sample of 149 stars that pass the selection criteria (green), and the metallicity distribution of the 46 stars that have a $\textrm{[Fe/H]}_{Pristine} \leq -3.0$ (blue). The percentages show the fraction of stars from the $\textrm{[Fe/H]}_{Pristine} \leq -3.0$ sample that are contained in the given metallicity ranges.}
    \label{FERRE_good_hist}
\end{figure}

Figure \ref{FERRE_good_hist} shows the FERRE metallicity distribution for the sample that satisfies the selection criteria. To visualize the success of the selection based on photometry, we plot in blue the metallicity distribution of the $\textrm{[Fe/H]}_{Pristine} \leq -3.0$ sample. The percentage of stars that fall in each region are shown, namely that 22\% of these stars still end up below $\textrm{[Fe/H]} \leq -3.0$, 50\% fall between $-3.0 < \textrm{[Fe/H]} \leq -2.5$, and 15\% are contaminants with $\textrm{[Fe/H]} \geq -2.0$. 

\begin{table*}
\centering
\caption{Numbers of stars with photometric predictions $\textrm{[Fe/H]}_{Pristine}$ below $-2.5$ and $-3.0$, the numbers of stars that are spectroscopically confirmed below those metallicities, and the success rates, given for the full spectroscopic sample, the sample after application of the selection criteria (described in Section \ref{sec:improve}), and the sample of stars with $\textrm{[Fe/H]}_{Pristine} \leq -3.0$.}
\label{success_rates}
\begin{tabular}{cccc}
\hline
\textbf{}                   & \textbf{Total observed} & \textbf{Selection criteria} & \textbf{$\textrm{[Fe/H]}_{Pristine} \leq -3.0$} \\ \hline
\textbf{Total number} & 205 & 149 & 46 \\
\textbf{$\textrm{[Fe/H]}_{Pristine} \leq -2.5$} & 163/205 (80\%) & 130/149 (87\%) & 46/46 (100\%) \\
\textbf{$\textrm{[Fe/H]}_{Pristine} \leq -3.0$} & 73/205 (36\%) & 46/149 (31\%) & 46/46 (100\%) \\
\hline
\textbf{$\textrm{[Fe/H]}_{\textrm{FERRE}} \leq -2.5$} & 119/205 (58\%) & 98/149 (66\%) & 33/46 (72\%) \\
\textbf{$\textrm{[Fe/H]}_{\textrm{FERRE}} \leq -3.0$} & 27/205 (13\%) & 25/149 (17\%) & 10/46 (22\%) \\
\hline
\textbf{$\textrm{[Fe/H]}_{\textrm{FERRE}} \geq -2.0$} & 42/205 (20\%) & 11/149 (7\%) & 7/46 (15\%) \\
\hline
\textbf{success $\textrm{[Fe/H]} \leq -2.5$} & 101/163 (62\%) & 91/130 (70\%) & - \\
\textbf{success $\textrm{[Fe/H]} \leq -3.0$} & 12/73 (16\%) & 10/46 (22\%) & 10/46 (22\%) \\
\hline
\end{tabular}
\end{table*}

Over the \emph{Pristine} footprint, covering  $\sim1000 \textrm{ deg}^2$ as of September 2016, we have photometrically identified 10~243 metal-poor star candidates with $V < 18$ that pass all of the selection criteria laid out in Section \ref{sec:improve}. The \emph{Pristine} survey does go deeper than this ($V \sim 20$), but this is the magnitude range accessible with $2-4$m class telescopes. The selected sample constitutes 1.3\% of all of the stars in the survey present in this magnitude range, and more than half of these have a predicted photometric metallicity $\textrm{[Fe/H]}_{Pristine} \leq -2.5$. Table \ref{Spectro_sample} summarizes the number of candidate stars split into magnitude ranges, where the first number given for each entry is the number of stars followed up and the second is the number of candidates in the full sample.

Although the success rates reported in this paper are based on a small sample of stars, we can still estimate the number of stars with $\textrm{[Fe/H]} \leq -3.0$ that we would expect to find in the entire $V < 18$ sample. However, since we have selected the best candidates available first, (i.e., we have observed a higher fraction of stars with $\textrm{[Fe/H]}_{Pristine} \leq -3.0$), we cannot directly scale the number of EMP stars found in our 205 star sub-sample to the number expected for the full sample. We therefore separate the sample into ranges of photometric metallicities, compute the relative fraction of EMP stars recovered in each metallicity range, and then scale these numbers to the total candidate sample. This calculation yields an expected number of $\sim 1000 - 1200$ EMP stars over the $\sim 1000 \textrm{ deg}^2$ \emph{Pristine} footprint (the final number is somewhat dependant on the bin size chosen for the metallicity ranges). Considering all observed stars, we therefore estimate a frequency of $\sim 1/800$ (1.25\%) for stars to have a metallicity of $\textrm{[Fe/H]} \leq -3.0$ for $14 < V < 18$ in the Galactic halo. 

Based on the Besan\c{c}on Model of stellar population synthesis of the Galaxy \citep{Robin03} -- for a similar sky region to the \emph{Pristine} footprint and a magnitude range of $14 < V < 18$ -- we expect a frequency of 1/2000 (0.05\%) for randomly selected halo stars to have $\textrm{[Fe/H]} \leq -3.0$. It should be noted that this is only a first order approximation, as the model relies on several assumptions about the metal-poor tail of the halo metallically distribution function. It should also be noted that our projections have been made based on a small sample of stars that preferentially occupy the brighter part of this magnitude range. However, as a coarse comparison, frequencies of expected EMP stars from simulated galaxy model predictions are in reasonable agreement with our observations.

\begin{table}
\centering
\caption{Number of candidate stars in different magnitude bins and metallicity ranges. The first number in each cell is the number of stars followed up with spectroscopy from the sample in this paper, and the second  is the total number of candidates as of September 2016 over the $\sim 1000 \textrm{ deg}^2$ \emph{Pristine} survey footprint. [Fe/H] values shown are photometric \emph{Pristine} $g-i$ metallicities.}
\label{Spectro_sample}
\begin{tabular}{cccc}
\hline
\textbf{}                   & \textbf{\# Candidates} & \textbf{{[}Fe/H{]}$\leq$-2.5} & \textbf{{[}Fe/H{]}$\leq$-3.0} \\ \hline
\textbf{$V < 15$}             & 47/213           & 30/166                           & 13/48                              \\
\textbf{$15 < V < 16$} & 114/797          & 91/554                            & 33/92                              \\
\textbf{$16 <V < 17$} & 29/2 388         & 28/1 549                           & 17/242                            \\
\textbf{$17 < V < 18$} & 15/6 845          & 14/4 354                           & 10/674                             \\ \hline
\textbf{Total }                      & 205/10 243        & 163/6 623                             & 73/1 056                            \\
\hline
\end{tabular}
\end{table}

Table \ref{comparison_table} summarizes the comparisons of the relative return for HES, SC14 and \emph{Pristine}. Other efforts have yielded similar or lower return rates as HES and SC14 \citep[e.g.,][]{AllendePrieto00}.

\renewcommand{\thefootnote}{\fnsymbol{footnote}}
\begin{table}
\centering
\caption{The relative fractions of metal-poor stars in \emph{Pristine} compared to other surveys.}
\label{comparison_table}
\begin{tabular}{cccc}
\hline
\textbf{Survey}             & \textbf{{[}Fe/H{]} \textless -3} & \textbf{{[}Fe/H{]} \textless -2.5} & \textbf{-3 \textless {[}Fe/H{]} \textless -2} \\ \hline
\textbf{Pristine}           & 17\%                             & 66\%                               & 76\%                                          \\
\textbf{HES}              & 4\%                            & 22\%\footnotemark[1]                              & 40\%\footnotemark[1]                                         \\
\textbf{SC14} & 3.8\%                            & -                                  & 32\%                           	\\
\hline              
\end{tabular}
\end{table}
\footnotetext[1]{These percentages are computed from the scaled sample presented in Table 3 of \citet{Schorck2009}.}

\subsection{Comparison to other surveys}

In order to compare these results to other surveys, we use the relative fractions of metal-poor stars from the selected sample of 149 stars. Although the success rates are a more telling quantification of the capabilities of \emph{Pristine} for finding EMP stars, it is more appropriate to use the relative fractions for a quantitative comparison to other works. This is because the \emph{Pristine} survey has the advantage over other metal-poor star searches that it can quantify the metallicity of its candidates, and select for example candidates with $\textrm{[Fe/H]}_{Pristine} \leq -3.0$, instead of labelling objects in a binary fashion as EMP candidates or not. 

SC14 report that $3.8^{+1.3}_{-1.1}\%$ of their candidate stars have an $\textrm{[Fe/H]} \lesssim -3.0$, and $32^{+3.0}_{-2.9}\%$ have $-3.0 \lesssim \textrm{[Fe/H]} \lesssim -2.0$, from high resolution follow-up of their selection with WISE and 2MASS magnitudes. Although we report significantly higher rates of 17\% and 76\%, respectively, it should be taken into consideration that they are using publicly available survey data and are specifically targeting bright stars, so they enjoy the advantage of large sky coverage and ease of spectroscopic follow-up. They also use near infrared magnitudes, which offers the advantage of being able to probe the crowded regions of the disk in the direction of the bulge \citep{Casey15}, although they are limited in the distance they can reach due to the bright nature of their sample.

The stellar content and metallicity distributions of HES are presented in \citet{Schorck2009}. In that paper, they report a fraction of stars with $\textrm{[Fe/H]} \leq -3.0$ of 7\% for their best-selected sample, and $3-4$\% for the other samples. Their best-selected sample totals 105 out of 1~638 stars, and constitutes only 6.4\% of their total accepted follow-up sample. This sample can be compared to the 22\% success rate of the best \emph{Pristine} sample, stars with an assigned photometric metallicity of $\textrm{[Fe/H]} \leq -3.0$. Taking the entire HES sample as a whole then yields 65 out of 1~638 stars with $\textrm{[Fe/H]} \leq -3.0$, a fraction of 4\%, and this can be compared to the relative fraction from the whole \emph{Pristine} sample of 17\%.

\subsection{Future follow-up strategy}

\citet{Schorck2009} report that for the bias-corrected HES metallicity distribution function, around 1-3\% of all $\textrm{[Fe/H]} \leq -3.0$ stars had a metallicity $\textrm{[Fe/H]} \leq -4.0$. \citet{Allende_Prieto_2014} report similar numbers for SDSS/BOSS, with 1 star at $\textrm{[Fe/H]} \leq -4.0$ out of 118 at $\textrm{[Fe/H]} \leq -3.0$ (see their Table 2). 

We can use these statistics to make projections of how many UMP stars we expect to find. Taking a conservative estimate, we expect one star with $\textrm{[Fe/H]} \leq -4.0$ for every $\sim 100$ stars with $\textrm{[Fe/H]} \leq -3.0$. We therefore predict that we will find $\sim 10 - 12$ UMP stars over the $\sim 1000\textrm{ deg}^2$ footprint in the magnitude range $V < 18$, given that we will uncover a projected $\sim 1000 - 1200$ stars with $\textrm{[Fe/H]} \leq -3.0$. Furthermore, it is not surprising that we have not yet found any UMP stars in our current sample of 205 stars (27 with $\textrm{[Fe/H]} \leq -3.0$). This sample is still quite small when compared to other surveys that have successfully found UMP stars, such as SDSS (with SEGUE and BOSS), and HES, which have both followed up many thousands of stars with low-resolution spectroscopy. 

Projecting forward into the future, with a larger footprint of $\sim 3000\textrm{ deg}^2$, we expect to find a statistical sample of several tens of these stars. Furthermore, if we can follow-up the fainter magnitude range of \emph{Pristine} ($18 < V < 20$), this not only would provide many more candidates, but also probe deeper into the halo and potentially result in the discovery of many more UMP stars.

Given the availability of time on $2-4$m class telescopes, it may be possible for our team to obtain a complete follow-up sample for the brighter magnitude ranges of our candidate sample, up to $V < 16$. For the magnitude ranges fainter than this, there are too many candidates to feasibly follow up with single slit spectrographs. However, this task would be well-suited to the upcoming new generations of multi-object spectrographs, such as WEAVE, 4MOST, PFS, and MSE.

\section{Conclusions}

Through an analysis of the first medium resolution spectroscopic sample from the follow-up programme of \emph{Pristine}, we have demonstrated that the narrow-band survey is very efficient at uncovering EMP stars in the Galactic halo. We used this sample to assess and refine the selection criteria for selecting photometric candidates for spectroscopic follow-up. This  included investigating whether infrared magnitudes from WISE and 2MASS could improve the selection efficiency, as was done by \citet{Schlaufman_Casey}, but this added information was only useful for the brightest Pristine stars ($V < 15$) and even then resulted in low completeness in the metal-poor regime of $\textrm{[Fe/H]} \leq -2.5$. Analyzing the selection criteria with a regularized regression technique, we confirmed that the $u$, $g$, $r$, $i$, and $CaHK$ magnitudes contain the most useful information for separating the sample by metallicity. 

The total spectroscopic sample consisted of 205 stars, of which 27 were found to have $\textrm{[Fe/H]} \leq -3.0$ and 119 were found to have $\textrm{[Fe/H]} \leq -2.5$. This sample was reduced to 149 stars by the refined photometric selection criteria, of which 25 had $\textrm{[Fe/H]} \leq -3.0$ (17\%) and 98 were found to have $\textrm{[Fe/H]} \leq -2.5$ (66\%). This return rate for finding EMP stars is unprecedented, with other surveys typically reporting values of 3-4\%. For stars predicted by \emph{Pristine} to be EMP with $\textrm{[Fe/H]} \leq -3.0$, we report a success rate of 22\% for confirming them as EMP, and for stars predicted to have a metallicity of $\textrm{[Fe/H]} \leq -2.5$ we report a success rate of 70\%. 

The \emph{Pristine} survey is ongoing, both with increasing sky coverage of the photometric footprint with CFHT/Megacam and with its spectroscopic follow-up campaign. Based on our statistics, we expect to uncover $\sim 1000 - 1200$ stars with $\textrm{[Fe/H]} \leq -3.0$ and $\sim 10 - 12$ stars with $\textrm{[Fe/H]} \leq -4.0$ per $1000\textrm{ deg}^2$ of survey area. In the future, we hope to expand our spectroscopic follow-up towards fainter magnitudes with the next generation of multi-object spectrographs. 

\section*{Acknowledgements}
We would like to thank the referee for their useful comments which helped to improve the paper. KY would also like to thank Gal Matijevi\v c and Jenn Wojno for their comments and insightful discussions. We gratefully thank the CFHT staff for performing the observations in queue mode, for their reactivity in adapting the schedule, and for answering our questions during the data-reduction process. We also thank the support astronomers and staff at the INT/WHT for their expertise and help with observations. We thank Nina Hernitschek for granting us access to the catalogue of PanSTARRS1 variability catalogue. ES and KY gratefully acknowledge funding by the Emmy Noether program from the Deutsche Forschungsgemeinschaft (DFG). NFM gratefully acknowledges funding from CNRS/INSU through the Programme National Galaxies et Cosmologie, and through the CNRS grant PICS07708. ES and KY benefited from the International Space Science Institute (ISSI) in Bern, CH, thanks to the funding of the Team ``The Formation and Evolution of the Galactic Halo''. DA acknowledges the Spanish Ministry of Economy and Competitiveness (MINECO) for the financial support received in the form of a Severo-Ochoa PhD fellowship, within the Severo-Ochoa International PhD Program. DA, CAP, and JIGH also acknowledge the Spanish ministry project MINECO AYA2014-56359-P. JIGH acknowledges financial support from the Spanish Ministry of Economy and Competitiveness (MINECO) under the 2013 Ram\'{o}n y Cajal program MINECO RYC-2013-14875. 

This paper is based on observations obtained with MegaPrime/MegaCam, a joint project of CFHT and CEA/DAPNIA, at the Canada-France-Hawaii Telescope (CFHT) which is operated by the National Research Council (NRC) of Canada, the Institut National des Science de l'Univers of the Centre National de la Recherche Scientifique (CNRS) of France, and the University of Hawaii.

Funding for the Sloan Digital Sky Survey IV has been provided by the Alfred P. Sloan Foundation, the U.S. Department of Energy Office of Science, and the Participating Institutions. SDSS acknowledges support and resources from the Center for High-Performance Computing at the University of Utah. The SDSS web site is www.sdss.org.

SDSS is managed by the Astrophysical Research Consortium for the Participating Institutions of the SDSS Collaboration including the Brazilian Participation Group, the Carnegie Institution for Science, Carnegie Mellon University, the Chilean Participation Group, the French Participation Group, Harvard-Smithsonian Center for Astrophysics, Instituto de Astrofísica de Canarias, The Johns Hopkins University, Kavli Institute for the Physics and Mathematics of the Universe (IPMU) / University of Tokyo, Lawrence Berkeley National Laboratory, Leibniz Institut für Astrophysik Potsdam (AIP), Max-Planck-Institut für Astronomie (MPIA Heidelberg), Max-Planck-Institut für Astrophysik (MPA Garching), Max-Planck-Institut für Extraterrestrische Physik (MPE), National Astronomical Observatories of China, New Mexico State University, New York University, University of Notre Dame, Observatório Nacional / MCTI, The Ohio State University, Pennsylvania State University, Shanghai Astronomical Observatory, United Kingdom Participation Group, Universidad Nacional Autónoma de México, University of Arizona, University of Colorado Boulder, University of Oxford, University of Portsmouth, University of Utah, University of Virginia, University of Washington, University of Wisconsin, Vanderbilt University, and Yale University.



\bibliographystyle{mn2e}
\bibliography{referee_reply_draft} 

\begin{thebibliography}{74}
\expandafter\ifx\csname natexlab\endcsname\relax\def\natexlab#1{#1}\fi

\bibitem[{{Aguado} {et~al}\mbox{.}(2016){Aguado}, {Allende Prieto},
  {Gonz{\'a}lez Hern{\'a}ndez}, {Carrera}, {Rebolo}, {Shetrone}, {Lambert}, \&
  {Fern{\'a}ndez-Alvar}}]{Aguado16}
{Aguado} D.~S., {Allende Prieto} C., {Gonz{\'a}lez Hern{\'a}ndez} J.~I.,
  {Carrera} R., {Rebolo} R., {Shetrone} M., {Lambert} D.~L.,
  {Fern{\'a}ndez-Alvar} E., 2016, \aap, 593, A10

\bibitem[{{Aguado} {et~al}\mbox{.}(2017{\natexlab{a}}){Aguado}, {Allende
  Prieto}, {Gonz{\'a}lez Hern{\'a}ndez}, {Rebolo}, \& {Caffau}}]{Aguado17a}
{Aguado} D.~S., {Allende Prieto} C., {Gonz{\'a}lez Hern{\'a}ndez} J.~I.,
  {Rebolo} R., {Caffau} E., 2017{\natexlab{a}}, \aap, 604, A9

\bibitem[{{Aguado} {et~al}\mbox{.}(2017{\natexlab{b}}){Aguado}, {Gonz{\'a}lez
  Hern{\'a}ndez}, {Allende Prieto}, \& {Rebolo}}]{Aguado17b}
{Aguado} D.~S., {Gonz{\'a}lez Hern{\'a}ndez} J.~I., {Allende Prieto} C.,
  {Rebolo} R., 2017{\natexlab{b}}, ArXiv e-prints

\bibitem[{{Allende Prieto} {et~al}\mbox{.}(2006){Allende Prieto}, {Beers},
  {Wilhelm}, {Newberg}, {Rockosi}, {Yanny}, \& {Lee}}]{Allende_Prieto_2006}
{Allende Prieto} C., {Beers} T.~C., {Wilhelm} R., {Newberg} H.~J., {Rockosi}
  C.~M., {Yanny} B., {Lee} Y.~S., 2006, \apj, 636, 804

\bibitem[{{Allende Prieto} {et~al}\mbox{.}(2015{\natexlab{a}}){Allende Prieto},
  {Fern{\'a}ndez-Alvar}, {Aguado}, {Gonz{\'a}lez Hern{\'a}ndez}, {Rebolo},
  {Lee}, {Beers}, {Rockosi}, \& {Ge}}]{AllendePrieto15a}
{Allende Prieto} C. {et~al.}, 2015{\natexlab{a}}, \aap, 579, A98

\bibitem[{{Allende Prieto} {et~al}\mbox{.}(2014){Allende Prieto},
  {Fern{\'a}ndez-Alvar}, {Schlesinger}, {Lee}, {Morrison}, {Schneider},
  {Beers}, {Bizyaev}, {Ebelke}, {Malanushenko}, {Malanushenko}, {Oravetz},
  {Pan}, {Simmons}, {Simmerer}, {Sobeck}, \& {Robin}}]{Allende_Prieto_2014}
---, 2014, \aap, 568, A7

\bibitem[{{Allende Prieto} {et~al}\mbox{.}(2000){Allende Prieto}, {Rebolo},
  {Garc{\'{\i}}a L{\'o}pez}, {Serra-Ricart}, {Beers}, {Rossi}, {Bonifacio}, \&
  {Molaro}}]{AllendePrieto00}
{Allende Prieto} C., {Rebolo} R., {Garc{\'{\i}}a L{\'o}pez} R.~J.,
  {Serra-Ricart} M., {Beers} T.~C., {Rossi} S., {Bonifacio} P., {Molaro} P.,
  2000, \aj, 120, 1516

\bibitem[{{Allende Prieto} {et~al}\mbox{.}(2015{\natexlab{b}}){Allende Prieto}
  {et~al.}}]{AllendePrieto15b}
{Allende Prieto} C., {et~al.}, 2015{\natexlab{b}}, in American Astronomical
  Society Meeting Abstracts, Vol. 225, American Astronomical Society Meeting
  Abstracts, p. 422.07

\bibitem[{{Alpher}, {Bethe} \& {Gamow}(1948){Alpher}, {Bethe}, \&
  {Gamow}}]{Alpher48}
{Alpher} R.~A., {Bethe} H., {Gamow} G., 1948, Physical Review, 73, 803

\bibitem[{{Alvarez} \& {Plez}(1998)}]{Alvarez_Plez_1998}
{Alvarez} R., {Plez} B., 1998, \aap, 330, 1109

\bibitem[{{An} {et~al}\mbox{.}(2013){An}, {Beers}, {Johnson}, {Pinsonneault},
  {Lee}, {Bovy}, {Ivezi{\'c}}, {Carollo}, \& {Newby}}]{An13}
{An} D. {et~al.}, 2013, \apj, 763, 65

\bibitem[{{An} {et~al}\mbox{.}(2015){An}, {Beers}, {Santucci}, {Carollo},
  {Placco}, {Lee}, \& {Rossi}}]{An15}
{An} D., {Beers} T.~C., {Santucci} R.~M., {Carollo} D., {Placco} V.~M., {Lee}
  Y.~S., {Rossi} S., 2015, \apjl, 813, L28

\bibitem[{{Anthony-Twarog} {et~al}\mbox{.}(2000){Anthony-Twarog}, {Sarajedini},
  {Twarog}, \& {Beers}}]{Twarog2000}
{Anthony-Twarog} B.~J., {Sarajedini} A., {Twarog} B.~A., {Beers} T.~C., 2000,
  \aj, 119, 2882

\bibitem[{{Anthony-Twarog} {et~al}\mbox{.}(1991){Anthony-Twarog}, {Twarog},
  {Laird}, \& {Payne}}]{Twarog91}
{Anthony-Twarog} B.~J., {Twarog} B.~A., {Laird} J.~B., {Payne} D., 1991, \aj,
  101, 1902

\bibitem[{{Aoki} {et~al}\mbox{.}(2013){Aoki}, {Beers}, {Lee}, {Honda}, {Ito},
  {Takada-Hidai}, {Frebel}, {Suda}, {Fujimoto}, {Carollo}, \&
  {Sivarani}}]{Aoki13}
{Aoki} W. {et~al.}, 2013, \aj, 145, 13

\bibitem[{{Beers} \& {Christlieb}(2005)}]{Beers05}
{Beers} T.~C., {Christlieb} N., 2005, \araa, 43, 531

\bibitem[{{Beers}, {Preston} \& {Shectman}(1985){Beers}, {Preston}, \&
  {Shectman}}]{Beers85}
{Beers} T.~C., {Preston} G.~W., {Shectman} S.~A., 1985, \aj, 90, 2089

\bibitem[{{Beers}, {Preston} \& {Shectman}(1992){Beers}, {Preston}, \&
  {Shectman}}]{Beers92}
---, 1992, \aj, 103, 1987

\bibitem[{{Beers} {et~al}\mbox{.}(1999){Beers}, {Rossi}, {Norris}, {Ryan}, \&
  {Shefler}}]{Beers99}
{Beers} T.~C., {Rossi} S., {Norris} J.~E., {Ryan} S.~G., {Shefler} T., 1999,
  \aj, 117, 981

\bibitem[{{Bessell} {et~al}\mbox{.}(2011){Bessell}, {Bloxham}, {Schmidt},
  {Keller}, {Tisserand}, \& {Francis}}]{Bessell2011}
{Bessell} M., {Bloxham} G., {Schmidt} B., {Keller} S., {Tisserand} P.,
  {Francis} P., 2011, \pasp, 123, 789

\bibitem[{{Bonifacio} {et~al}\mbox{.}(2009){Bonifacio}, {Spite}, {Cayrel},
  {Hill}, {Spite}, {Fran{\c c}ois}, {Plez}, {Ludwig}, {Caffau}, {Molaro},
  {Depagne}, {Andersen}, {Barbuy}, {Beers}, {Nordstr{\"o}m}, \&
  {Primas}}]{Bonifacio_2009}
{Bonifacio} P. {et~al.}, 2009, \aap, 501, 519

\bibitem[{{Burbidge} {et~al}\mbox{.}(1957){Burbidge}, {Burbidge}, {Fowler}, \&
  {Hoyle}}]{Burbidge57}
{Burbidge} E.~M., {Burbidge} G.~R., {Fowler} W.~A., {Hoyle} F., 1957, Reviews
  of Modern Physics, 29, 547

\bibitem[{{Caffau} {et~al}\mbox{.}(2013){Caffau}, {Bonifacio}, {Sbordone},
  {Fran{\c c}ois}, {Monaco}, {Spite}, {Plez}, {Cayrel}, {Christlieb}, {Clark},
  {Glover}, {Klessen}, {Koch}, {Ludwig}, {Spite}, {Steffen}, \&
  {Zaggia}}]{Caffau13}
{Caffau} E. {et~al.}, 2013, \aap, 560, A71

\bibitem[{{Caffau} {et~al}\mbox{.}(2017){Caffau}, {Bonifacio}, {Starkenburg},
  {Martin}, {Youakim}, {Henden}, {Gonzalez Hernandez}, {Aguado}, {Allende
  Prieto}, {Venn}, \& {Jablonka}}]{Caffau17}
---, 2017, ArXiv e-prints

\bibitem[{{Carollo} {et~al}\mbox{.}(2012){Carollo}, {Beers}, {Bovy},
  {Sivarani}, {Norris}, {Freeman}, {Aoki}, {Lee}, \& {Kennedy}}]{Carollo12}
{Carollo} D. {et~al.}, 2012, \apj, 744, 195

\bibitem[{{Carollo} {et~al}\mbox{.}(2010){Carollo}, {Beers}, {Chiba}, {Norris},
  {Freeman}, {Lee}, {Ivezi{\'c}}, {Rockosi}, \& {Yanny}}]{Carollo10}
---, 2010, \apj, 712, 692

\bibitem[{{Casey} \& {Schlaufman}(2015)}]{Casey15}
{Casey} A.~R., {Schlaufman} K.~C., 2015, \apj, 809, 110

\bibitem[{{Cayrel} {et~al}\mbox{.}(2004){Cayrel}, {Depagne}, {Spite}, {Hill},
  {Spite}, {Fran{\c c}ois}, {Plez}, {Beers}, {Primas}, {Andersen}, {Barbuy},
  {Bonifacio}, {Molaro}, \& {Nordstr{\"o}m}}]{Cayrel_2004}
{Cayrel} R. {et~al.}, 2004, \aap, 416, 1117

\bibitem[{{Christlieb}, {Wisotzki} \& {Gra{\ss}hoff}(2002){Christlieb},
  {Wisotzki}, \& {Gra{\ss}hoff}}]{Christlieb02}
{Christlieb} N., {Wisotzki} L., {Gra{\ss}hoff} G., 2002, \aap, 391, 397

\bibitem[{{Dalton} {et~al}\mbox{.}(2016){Dalton}, {Trager}, {Abrams},
  {Bonifacio}, {Aguerri}, {Middleton}, {Benn}, {Dee}, {Say{\`e}de}, {Lewis},
  {Pragt}, {Pico}, {Walton}, {Rey}, {Allende Prieto}, {Pe{\~n}ate}, {Lhome},
  {Ag{\'o}cs}, {Alonso}, {Terrett}, {Brock}, {Gilbert}, {Schallig}, {Ridings},
  {Guinouard}, {Verheijen}, {Tosh}, {Rogers}, {Lee}, {Steele}, {Stuik},
  {Tromp}, {Jask{\'o}}, {Carrasco}, {Farcas}, {Kragt}, {Lesman}, {Kroes},
  {Mottram}, {Bates}, {Rodriguez}, {Gribbin}, {Delgado}, {Herreros}, {Martin},
  {Cano}, {Navarro}, {Irwin}, {Lewis}, {Gonzalez Solares}, {Murphy}, {Worley},
  {Bassom}, {O'Mahoney}, {Bianco}, {Zurita}, {ter Horst}, {Molinari}, {Lodi},
  {Guerra}, {Martin}, {Vallenari}, {Salasnich}, {Baruffolo}, {Jin}, {Hill},
  {Smith}, {Drew}, {Poggianti}, {Pieri}, {Dominquez Palmero}, \&
  {Farina}}]{Dalton_2016}
{Dalton} G. {et~al.}, 2016, in \procspie, Vol. 9908, Society of Photo-Optical
  Instrumentation Engineers (SPIE) Conference Series, p. 99081G

\bibitem[{{Dalton} {et~al}\mbox{.}(2014){Dalton}, {Trager}, {Abrams},
  {Bonifacio}, {L{\'o}pez Aguerri}, {Middleton}, {Benn}, {Dee}, {Say{\`e}de},
  {Lewis}, {Pragt}, {Pico}, {Walton}, {Rey}, {Allende Prieto}, {Pe{\~n}ate},
  {Lhome}, {Ag{\'o}cs}, {Alonso}, {Terrett}, {Brock}, {Gilbert}, {Ridings},
  {Guinouard}, {Verheijen}, {Tosh}, {Rogers}, {Steele}, {Stuik}, {Tromp},
  {Jasko}, {Kragt}, {Lesman}, {Mottram}, {Bates}, {Gribbin}, {Rodriguez},
  {Delgado}, {Martin}, {Cano}, {Navarro}, {Irwin}, {Lewis}, {Gonzalez Solares},
  {O'Mahony}, {Bianco}, {Zurita}, {ter Horst}, {Molinari}, {Lodi}, {Guerra},
  {Vallenari}, \& {Baruffolo}}]{Dalton14}
---, 2014, in \procspie, Vol. 9147, Ground-based and Airborne Instrumentation
  for Astronomy V, p. 91470L

\bibitem[{{Dalton} {et~al}\mbox{.}(2012){Dalton}, {Trager}, {Abrams}, {Carter},
  {Bonifacio}, {Aguerri}, {MacIntosh}, {Evans}, {Lewis}, {Navarro}, {Agocs},
  {Dee}, {Rousset}, {Tosh}, {Middleton}, {Pragt}, {Terrett}, {Brock}, {Benn},
  {Verheijen}, {Cano Infantes}, {Bevil}, {Steele}, {Mottram}, {Bates},
  {Gribbin}, {Rey}, {Rodriguez}, {Delgado}, {Guinouard}, {Walton}, {Irwin},
  {Jagourel}, {Stuik}, {Gerlofsma}, {Roelfsma}, {Skillen}, {Ridings},
  {Balcells}, {Daban}, {Gouvret}, {Venema}, \& {Girard}}]{Dalton12}
---, 2012, in \procspie, Vol. 8446, Ground-based and Airborne Instrumentation
  for Astronomy IV, p. 84460P

\bibitem[{{Dawson} {et~al}\mbox{.}(2013){Dawson}, {Schlegel}, {Ahn},
  {Anderson}, {Aubourg}, {Bailey}, {Barkhouser}, {Bautista}, {Beifiori},
  {Berlind}, {Bhardwaj}, {Bizyaev}, {Blake}, {Blanton}, {Blomqvist}, {Bolton},
  {Borde}, {Bovy}, {Brandt}, {Brewington}, {Brinkmann}, {Brown}, {Brownstein},
  {Bundy}, {Busca}, {Carithers}, {Carnero}, {Carr}, {Chen}, {Comparat},
  {Connolly}, {Cope}, {Croft}, {Cuesta}, {da Costa}, {Davenport}, {Delubac},
  {de Putter}, {Dhital}, {Ealet}, {Ebelke}, {Eisenstein}, {Escoffier}, {Fan},
  {Filiz Ak}, {Finley}, {Font-Ribera}, {G{\'e}nova-Santos}, {Gunn}, {Guo},
  {Haggard}, {Hall}, {Hamilton}, {Harris}, {Harris}, {Ho}, {Hogg}, {Holder},
  {Honscheid}, {Huehnerhoff}, {Jordan}, {Jordan}, {Kauffmann}, {Kazin},
  {Kirkby}, {Klaene}, {Kneib}, {Le Goff}, {Lee}, {Long}, {Loomis}, {Lundgren},
  {Lupton}, {Maia}, {Makler}, {Malanushenko}, {Malanushenko}, {Mandelbaum},
  {Manera}, {Maraston}, {Margala}, {Masters}, {McBride}, {McDonald}, {McGreer},
  {McMahon}, {Mena}, {Miralda-Escud{\'e}}, {Montero-Dorta}, {Montesano},
  {Muna}, {Myers}, {Naugle}, {Nichol}, {Noterdaeme}, {Nuza}, {Olmstead},
  {Oravetz}, {Oravetz}, {Owen}, {Padmanabhan}, {Palanque-Delabrouille}, {Pan},
  {Parejko}, {P{\^a}ris}, {Percival}, {P{\'e}rez-Fournon},
  {P{\'e}rez-R{\`a}fols}, {Petitjean}, {Pfaffenberger}, {Pforr}, {Pieri},
  {Prada}, {Price-Whelan}, {Raddick}, {Rebolo}, {Rich}, {Richards}, {Rockosi},
  {Roe}, {Ross}, {Ross}, {Rossi}, {Rubi{\~n}o-Martin}, {Samushia},
  {S{\'a}nchez}, {Sayres}, {Schmidt}, {Schneider}, {Sc{\'o}ccola}, {Seo},
  {Shelden}, {Sheldon}, {Shen}, {Shu}, {Slosar}, {Smee}, {Snedden}, {Stauffer},
  {Steele}, {Strauss}, {Streblyanska}, {Suzuki}, {Swanson}, {Tal}, {Tanaka},
  {Thomas}, {Tinker}, {Tojeiro}, {Tremonti}, {Vargas Maga{\~n}a}, {Verde},
  {Viel}, {Wake}, {Watson}, {Weaver}, {Weinberg}, {Weiner}, {West}, {White},
  {Wood-Vasey}, {Yeche}, {Zehavi}, {Zhao}, \& {Zheng}}]{Dawson_2013}
{Dawson} K.~S. {et~al.}, 2013, \aj, 145, 10

\bibitem[{{de Jong} {et~al}\mbox{.}(2016){de Jong}, {Barden}, {Bellido-Tirado},
  {Brynnel}, {Frey}, {Giannone}, {Haynes}, {Johl}, {Phillips}, {Schnurr},
  {Walcher}, {Winkler}, {Ansorge}, {Feltzing}, {McMahon}, {Baker}, {Caillier},
  {Dwelly}, {Gaessler}, {Iwert}, {Mandel}, {Piskunov}, {Pragt}, {Walton},
  {Bensby}, {Bergemann}, {Chiappini}, {Christlieb}, {Cioni}, {Driver},
  {Finoguenov}, {Helmi}, {Irwin}, {Kitaura}, {Kneib}, {Liske}, {Merloni},
  {Minchev}, {Richard}, \& {Starkenburg}}]{deJong16}
{de Jong} R.~S. {et~al.}, 2016, in \procspie, Vol. 9908, Society of
  Photo-Optical Instrumentation Engineers (SPIE) Conference Series, p. 99081O

\bibitem[{Efron {et~al}\mbox{.}(2004)Efron, Hastie, Johnstone, \&
  Tibshirani}]{efron2004}
Efron B., Hastie T., Johnstone I., Tibshirani R., 2004, The Annals of
  statistics, 32, 407

\bibitem[{{Eisenstein} {et~al}\mbox{.}(2011){Eisenstein}, {Weinberg}, {Agol},
  {Aihara}, {Allende Prieto}, {Anderson}, {Arns}, {Aubourg}, {Bailey},
  {Balbinot}, \& et~al.}]{Eisenstein_2011}
{Eisenstein} D.~J. {et~al.}, 2011, \aj, 142, 72

\bibitem[{{Fern{\'a}ndez-Alvar} {et~al}\mbox{.}(2016){Fern{\'a}ndez-Alvar},
  {Allende Prieto}, {Beers}, {Lee}, {Masseron}, \&
  {Schneider}}]{FernandezAlvar16}
{Fern{\'a}ndez-Alvar} E., {Allende Prieto} C., {Beers} T.~C., {Lee} Y.~S.,
  {Masseron} T., {Schneider} D.~P., 2016, \aap, 593, A28

\bibitem[{{Fern{\'a}ndez-Alvar} {et~al}\mbox{.}(2015){Fern{\'a}ndez-Alvar},
  {Allende Prieto}, {Schlesinger}, {Beers}, {Robin}, {Schneider}, {Lee},
  {Bizyaev}, {Ebelke}, {Malanushenko}, {Malanushenko}, {Oravetz}, {Pan}, \&
  {Simmons}}]{FernandezAlvar15}
{Fern{\'a}ndez-Alvar} E. {et~al.}, 2015, \aap, 577, A81

\bibitem[{{Frebel} {et~al}\mbox{.}(2006){Frebel}, {Christlieb}, {Norris},
  {Beers}, {Bessell}, {Rhee}, {Fechner}, {Marsteller}, {Rossi}, {Thom},
  {Wisotzki}, \& {Reimers}}]{Frebel06}
{Frebel} A. {et~al.}, 2006, \apj, 652, 1585

\bibitem[{{Frebel} \& {Norris}(2015)}]{Frebel15}
{Frebel} A., {Norris} J.~E., 2015, \araa, 53, 631

\bibitem[{{Freeman} \& {Bland-Hawthorn}(2002)}]{Freeman02}
{Freeman} K., {Bland-Hawthorn} J., 2002, \araa, 40, 487

\bibitem[{{Gustafsson} {et~al}\mbox{.}(2008){Gustafsson}, {Edvardsson},
  {Eriksson}, {J{\o}rgensen}, {Nordlund}, \& {Plez}}]{Gustafsson_2008}
{Gustafsson} B., {Edvardsson} B., {Eriksson} K., {J{\o}rgensen} U.~G.,
  {Nordlund} {\AA}., {Plez} B., 2008, \aap, 486, 951

\bibitem[{{Henden} \& {Munari}(2014)}]{Henden_2014}
{Henden} A., {Munari} U., 2014, Contributions of the Astronomical Observatory
  Skalnate Pleso, 43, 518

\bibitem[{{Henden} {et~al}\mbox{.}(2015){Henden}, {Levine}, {Terrell}, \&
  {Welch}}]{Henden_2015}
{Henden} A.~A., {Levine} S., {Terrell} D., {Welch} D.~L., 2015, in American
  Astronomical Society Meeting Abstracts, Vol. 225, American Astronomical
  Society Meeting Abstracts, p. 336.16

\bibitem[{{Henden} {et~al}\mbox{.}(2009){Henden}, {Welch}, {Terrell}, \&
  {Levine}}]{Henden_2009}
{Henden} A.~A., {Welch} D.~L., {Terrell} D., {Levine} S.~E., 2009, in American
  Astronomical Society Meeting Abstracts, Vol. 214, American Astronomical
  Society Meeting Abstracts \#214, p. 669

\bibitem[{{Hernitschek} {et~al}\mbox{.}(2016){Hernitschek}, {Schlafly},
  {Sesar}, {Rix}, {Hogg}, {Ivezi{\'c}}, {Grebel}, {Bell}, {Martin}, {Burgett},
  {Flewelling}, {Hodapp}, {Kaiser}, {Magnier}, {Metcalfe}, {Wainscoat}, \&
  {Waters}}]{Paanstars_variable}
{Hernitschek} N. {et~al.}, 2016, \apj, 817, 73

\bibitem[{{Howes} {et~al}\mbox{.}(2016){Howes}, {Asplund}, {Keller}, {Casey},
  {Yong}, {Lind}, {Frebel}, {Hays}, {Alves-Brito}, {Bessell}, {Casagrande},
  {Marino}, {Nataf}, {Owen}, {Da Costa}, {Schmidt}, \& {Tisserand}}]{Howes16}
{Howes} L.~M. {et~al.}, 2016, \mnras, 460, 884

\bibitem[{{Howes} {et~al}\mbox{.}(2015){Howes}, {Casey}, {Asplund}, {Keller},
  {Yong}, {Nataf}, {Poleski}, {Lind}, {Kobayashi}, {Owen}, {Ness}, {Bessell},
  {da Costa}, {Schmidt}, {Tisserand}, {Udalski}, {Szyma{\'n}ski},
  {Soszy{\'n}ski}, {Pietrzy{\'n}ski}, {Ulaczyk}, {Wyrzykowski}, {Pietrukowicz},
  {Skowron}, {Koz{\l}owski}, \& {Mr{\'o}z}}]{Howes15}
---, 2015, \nat, 527, 484

\bibitem[{{Ibata} {et~al}\mbox{.}(2014){Ibata}, {Lewis}, {McConnachie},
  {Martin}, {Irwin}, {Ferguson}, {Babul}, {Bernard}, {Chapman}, {Collins},
  {Fardal}, {Mackey}, {Navarro}, {Pe{\~n}arrubia}, {Rich}, {Tanvir}, \&
  {Widrow}}]{Ibata_2014}
{Ibata} R.~A. {et~al.}, 2014, \apj, 780, 128

\bibitem[{{Irwin} \& {Lewis}(2001)}]{Irwin_2001}
{Irwin} M., {Lewis} J., 2001, \nar, 45, 105

\bibitem[{{Ivezi{\'c}} {et~al}\mbox{.}(2008){Ivezi{\'c}}, {Sesar}, {Juri{\'c}},
  {Bond}, {Dalcanton}, {Rockosi}, {Yanny}, {Newberg}, {Beers}, {Allende
  Prieto}, {Wilhelm}, {Lee}, {Sivarani}, {Norris}, {Bailer-Jones}, {Re
  Fiorentin}, {Schlegel}, {Uomoto}, {Lupton}, {Knapp}, {Gunn}, {Covey}, {Allyn
  Smith}, {Miknaitis}, {Doi}, {Tanaka}, {Fukugita}, {Kent}, {Finkbeiner},
  {Munn}, {Pier}, {Quinn}, {Hawley}, {Anderson}, {Kiuchi}, {Chen}, {Bushong},
  {Sohi}, {Haggard}, {Kimball}, {Barentine}, {Brewington}, {Harvanek},
  {Kleinman}, {Krzesinski}, {Long}, {Nitta}, {Snedden}, {Lee}, {Harris},
  {Brinkmann}, {Schneider}, \& {York}}]{Ivezic08}
{Ivezi{\'c}} {\v Z}. {et~al.}, 2008, \apj, 684, 287

\bibitem[{{Keller} {et~al}\mbox{.}(2007){Keller}, {Schmidt}, {Bessell},
  {Conroy}, {Francis}, {Granlund}, {Kowald}, {Oates}, {Martin-Jones},
  {Preston}, {Tisserand}, {Vaccarella}, \& {Waterson}}]{Keller07}
{Keller} S.~C. {et~al.}, 2007, \pasa, 24, 1

\bibitem[{{Koch} {et~al}\mbox{.}(2016){Koch}, {McWilliam}, {Preston}, \&
  {Thompson}}]{Koch16}
{Koch} A., {McWilliam} A., {Preston} G.~W., {Thompson} I.~B., 2016, \aap, 587,
  A124

\bibitem[{{Lamb} {et~al}\mbox{.}(2017){Lamb}, {Venn}, {Andersen}, {Oya},
  {Shetrone}, {Fattahi}, {Howes}, {Asplund}, {Lardi{\`e}re}, {Akiyama}, {Ono},
  {Terada}, {Hayano}, {Suzuki}, {Blain}, {Jackson}, {Correia}, {Youakim}, \&
  {Bradley}}]{Lamb17}
{Lamb} M. {et~al.}, 2017, \mnras, 465, 3536

\bibitem[{{Lokhorst} {et~al}\mbox{.}(2016){Lokhorst}, {Starkenburg},
  {McConnachie}, {Navarro}, {Ferrarese}, {C{\^o}t{\'e}}, {Liu}, {Peng}, {Gwyn},
  {Cuillandre}, \& {Guhathakurta}}]{Lokhorst}
{Lokhorst} D. {et~al.}, 2016, \apj, 819, 124

\bibitem[{{McConnachie} {et~al}\mbox{.}(2016){McConnachie}, {Babusiaux},
  {Balogh}, {Caffau}, {C{\^o}t{\'e}}, {Driver}, {Robotham}, {Starkenburg},
  {Venn}, {Walker}, {Bauman}, {Flagey}, {Ho}, {Isani}, {Laychak}, {Mignot},
  {Murowinski}, {Salmon}, {Simons}, {Szeto}, {Vermeulen}, \&
  {Withington}}]{McConnachie16}
{McConnachie} A.~W. {et~al.}, 2016, ArXiv e-prints

\bibitem[{{Plez}(2008)}]{Plez_2008}
{Plez} B., 2008, Physica Scripta Volume T, 133, 014003

\bibitem[{{Robin} {et~al}\mbox{.}(2003){Robin}, {Reyl{\'e}}, {Derri{\`e}re}, \&
  {Picaud}}]{Robin03}
{Robin} A.~C., {Reyl{\'e}} C., {Derri{\`e}re} S., {Picaud} S., 2003, \aap, 409,
  523

\bibitem[{{Sandage} \& {Eggen}(1959)}]{Sandage_Eggen59}
{Sandage} A.~R., {Eggen} O.~J., 1959, \mnras, 119, 278

\bibitem[{{Schlaufman} \& {Casey}(2014)}]{Schlaufman_Casey}
{Schlaufman} K.~C., {Casey} A.~R., 2014, \apj, 797, 13

\bibitem[{{Sch{\"o}rck} {et~al}\mbox{.}(2009){Sch{\"o}rck}, {Christlieb},
  {Cohen}, {Beers}, {Shectman}, {Thompson}, {McWilliam}, {Bessell}, {Norris},
  {Mel{\'e}ndez}, {Ram{\'{\i}}rez}, {Haynes}, {Cass}, {Hartley}, {Russell},
  {Watson}, {Zickgraf}, {Behnke}, {Fechner}, {Fuhrmeister}, {Barklem},
  {Edvardsson}, {Frebel}, {Wisotzki}, \& {Reimers}}]{Schorck2009}
{Sch{\"o}rck} T. {et~al.}, 2009, \aap, 507, 817

\bibitem[{{Schwarzschild}, {Searle} \& {Howard}(1955){Schwarzschild}, {Searle},
  \& {Howard}}]{Schwarzschild55}
{Schwarzschild} M., {Searle} L., {Howard} R., 1955, \apj, 122, 353

\bibitem[{{Skrutskie} {et~al}\mbox{.}(2006){Skrutskie}, {Cutri}, {Stiening},
  {Weinberg}, {Schneider}, {Carpenter}, {Beichman}, {Capps}, {Chester},
  {Elias}, {Huchra}, {Liebert}, {Lonsdale}, {Monet}, {Price}, {Seitzer},
  {Jarrett}, {Kirkpatrick}, {Gizis}, {Howard}, {Evans}, {Fowler}, {Fullmer},
  {Hurt}, {Light}, {Kopan}, {Marsh}, {McCallon}, {Tam}, {Van Dyk}, \&
  {Wheelock}}]{Skrutskie06}
{Skrutskie} M.~F. {et~al.}, 2006, \aj, 131, 1163

\bibitem[{{Sk{\'u}lad{\'o}ttir} {et~al}\mbox{.}(2015){Sk{\'u}lad{\'o}ttir},
  {Tolstoy}, {Salvadori}, {Hill}, {Pettini}, {Shetrone}, \&
  {Starkenburg}}]{Skuladottir15}
{Sk{\'u}lad{\'o}ttir} {\'A}., {Tolstoy} E., {Salvadori} S., {Hill} V.,
  {Pettini} M., {Shetrone} M.~D., {Starkenburg} E., 2015, \aap, 574, A129

\bibitem[{{Starkenburg} {et~al}\mbox{.}(2013){Starkenburg}, {Hill}, {Tolstoy},
  {Fran{\c c}ois}, {Irwin}, {Boschman}, {Venn}, {de Boer}, {Lemasle},
  {Jablonka}, {Battaglia}, {Groot}, \& {Kaper}}]{Starkenburg13}
{Starkenburg} E. {et~al.}, 2013, \aap, 549, A88

\bibitem[{{Starkenburg} {et~al}\mbox{.}(2017a){Starkenburg}, {Martin},
  {Youakim}, {Aguado}, {Allende Prieto}, {Arentsen}, {Bernard}, {Bonifacio},
  {Caffau}, {Carlberg}, {Cote}, {Fouesneau}, {Francois}, {Franke}, {Gonzalez
  Hernandez}, {Gwyn}, {Hill}, {Ibata}, {Jablonka}, {Longeard}, {McConnachie},
  {Navarro}, {Sanchez-Janssen}, {Tolstoy}, \& {Venn}}]{starkenburg17a}
---, 2017a, ArXiv e-prints

\bibitem[{{Starkenburg} {et~al}\mbox{.}(2017b){Starkenburg}, {Oman}, {Navarro},
  {Crain}, {Fattahi}, {Frenk}, {Sawala}, \& {Schaye}}]{Starkenburg17b}
{Starkenburg} E., {Oman} K.~A., {Navarro} J.~F., {Crain} R.~A., {Fattahi} A.,
  {Frenk} C.~S., {Sawala} T., {Schaye} J., 2017b, \mnras, 465, 2212

\bibitem[{{Takada} {et~al}\mbox{.}(2014){Takada}, {Ellis}, {Chiba}, {Greene},
  {Aihara}, {Arimoto}, {Bundy}, {Cohen}, {Dor{\'e}}, {Graves}, {Gunn},
  {Heckman}, {Hirata}, {Ho}, {Kneib}, {F{\`e}vre}, {Lin}, {More}, {Murayama},
  {Nagao}, {Ouchi}, {Seiffert}, {Silverman}, {Sodr{\'e}}, {Spergel}, {Strauss},
  {Sugai}, {Suto}, {Takami}, \& {Wyse}}]{Takada14}
{Takada} M. {et~al.}, 2014, \pasj, 66, R1

\bibitem[{Tibshirani(1996)}]{tibshirani1996}
Tibshirani R., 1996, Journal of the Royal Statistical Society. Series B
  (Methodological), 267

\bibitem[{{Tody}(1986)}]{Tody_IRAF}
{Tody} D., 1986, in \procspie, Vol. 627, Instrumentation in astronomy VI,
  {Crawford} D.~L., ed., p. 733

\bibitem[{{Wallerstein}(1962)}]{Wallerstein62}
{Wallerstein} G., 1962, \apjs, 6, 407

\bibitem[{{Wright} {et~al}\mbox{.}(2010){Wright}, {Eisenhardt}, {Mainzer},
  {Ressler}, {Cutri}, {Jarrett}, {Kirkpatrick}, {Padgett}, {McMillan},
  {Skrutskie}, {Stanford}, {Cohen}, {Walker}, {Mather}, {Leisawitz}, {Gautier},
  {McLean}, {Benford}, {Lonsdale}, {Blain}, {Mendez}, {Irace}, {Duval}, {Liu},
  {Royer}, {Heinrichsen}, {Howard}, {Shannon}, {Kendall}, {Walsh}, {Larsen},
  {Cardon}, {Schick}, {Schwalm}, {Abid}, {Fabinsky}, {Naes}, \& {Tsai}}]{WISE}
{Wright} E.~L. {et~al.}, 2010, \aj, 140, 1868

\bibitem[{{Yanny} {et~al}\mbox{.}(2009){Yanny}, {Rockosi}, {Newberg}, {Knapp},
  {Adelman-McCarthy}, {Alcorn}, {Allam}, {Allende Prieto}, {An}, {Anderson},
  {Anderson}, {Bailer-Jones}, {Bastian}, {Beers}, {Bell}, {Belokurov},
  {Bizyaev}, {Blythe}, {Bochanski}, {Boroski}, {Brinchmann}, {Brinkmann},
  {Brewington}, {Carey}, {Cudworth}, {Evans}, {Evans}, {Gates}, {G{\"a}nsicke},
  {Gillespie}, {Gilmore}, {Nebot Gomez-Moran}, {Grebel}, {Greenwell}, {Gunn},
  {Jordan}, {Jordan}, {Harding}, {Harris}, {Hendry}, {Holder}, {Ivans},
  {Ivezi{\v c}}, {Jester}, {Johnson}, {Kent}, {Kleinman}, {Kniazev},
  {Krzesinski}, {Kron}, {Kuropatkin}, {Lebedeva}, {Lee}, {French Leger},
  {L{\'e}pine}, {Levine}, {Lin}, {Long}, {Loomis}, {Lupton}, {Malanushenko},
  {Malanushenko}, {Margon}, {Martinez-Delgado}, {McGehee}, {Monet}, {Morrison},
  {Munn}, {Neilsen}, {Nitta}, {Norris}, {Oravetz}, {Owen}, {Padmanabhan},
  {Pan}, {Peterson}, {Pier}, {Platson}, {Re Fiorentin}, {Richards}, {Rix},
  {Schlegel}, {Schneider}, {Schreiber}, {Schwope}, {Sibley}, {Simmons},
  {Snedden}, {Allyn Smith}, {Stark}, {Stauffer}, {Steinmetz}, {Stoughton},
  {SubbaRao}, {Szalay}, {Szkody}, {Thakar}, {Sivarani}, {Tucker}, {Uomoto},
  {Vanden Berk}, {Vidrih}, {Wadadekar}, {Watters}, {Wilhelm}, {Wyse}, {Yarger},
  \& {Zucker}}]{Yanny_SEGUE}
{Yanny} B. {et~al.}, 2009, \aj, 137, 4377

\bibitem[{{York} {et~al}\mbox{.}(2000){York}, {Adelman}, {Anderson},
  {Anderson}, {Annis}, {Bahcall}, {Bakken}, {Barkhouser}, {Bastian}, {Berman},
  {Boroski}, {Bracker}, {Briegel}, {Briggs}, {Brinkmann}, {Brunner}, {Burles},
  {Carey}, {Carr}, {Castander}, {Chen}, {Colestock}, {Connolly}, {Crocker},
  {Csabai}, {Czarapata}, {Davis}, {Doi}, {Dombeck}, {Eisenstein}, {Ellman},
  {Elms}, {Evans}, {Fan}, {Federwitz}, {Fiscelli}, {Friedman}, {Frieman},
  {Fukugita}, {Gillespie}, {Gunn}, {Gurbani}, {de Haas}, {Haldeman}, {Harris},
  {Hayes}, {Heckman}, {Hennessy}, {Hindsley}, {Holm}, {Holmgren}, {Huang},
  {Hull}, {Husby}, {Ichikawa}, {Ichikawa}, {Ivezi{\'c}}, {Kent}, {Kim},
  {Kinney}, {Klaene}, {Kleinman}, {Kleinman}, {Knapp}, {Korienek}, {Kron},
  {Kunszt}, {Lamb}, {Lee}, {Leger}, {Limmongkol}, {Lindenmeyer}, {Long},
  {Loomis}, {Loveday}, {Lucinio}, {Lupton}, {MacKinnon}, {Mannery}, {Mantsch},
  {Margon}, {McGehee}, {McKay}, {Meiksin}, {Merelli}, {Monet}, {Munn},
  {Narayanan}, {Nash}, {Neilsen}, {Neswold}, {Newberg}, {Nichol}, {Nicinski},
  {Nonino}, {Okada}, {Okamura}, {Ostriker}, {Owen}, {Pauls}, {Peoples},
  {Peterson}, {Petravick}, {Pier}, {Pope}, {Pordes}, {Prosapio},
  {Rechenmacher}, {Quinn}, {Richards}, {Richmond}, {Rivetta}, {Rockosi},
  {Ruthmansdorfer}, {Sandford}, {Schlegel}, {Schneider}, {Sekiguchi}, {Sergey},
  {Shimasaku}, {Siegmund}, {Smee}, {Smith}, {Snedden}, {Stone}, {Stoughton},
  {Strauss}, {Stubbs}, {SubbaRao}, {Szalay}, {Szapudi}, {Szokoly}, {Thakar},
  {Tremonti}, {Tucker}, {Uomoto}, {Vanden Berk}, {Vogeley}, {Waddell}, {Wang},
  {Watanabe}, {Weinberg}, {Yanny}, {Yasuda}, \& {SDSS
  Collaboration}}]{York_2000}
{York} D.~G. {et~al.}, 2000, \aj, 120, 1579

\end{thebibliography}

\bsp	

\end{document}